\documentclass[12pt,preprint]{aastex}
\usepackage {graphicx}
\usepackage{aastexug}

\shorttitle{Magnetic coupling of a BH with its disk}
\shortauthors{D. X. Wang et al.}

\begin{document}

\title{Magnetic coupling of a rotating black hole \\
 with its surrounding accretion disk}

\author{Ding-Xiong Wang\altaffilmark{1},  Ren-Yi Ma  and  Wei-Hua Lei}
\affil{Department of Physics, Huazhong University of Science and
Technology,Wuhan, 430074, P. R. China} \and
\author{Guo-Zheng Yao}
\affil{Department of Physics, Beijing Normal University, Beijing,
100875, P. R. China}

\altaffiltext{1}{Send offprint requests to: D. X. Wang
(dxwang@hust.edu.cn)}

\begin{abstract}

Effects of magnetic coupling (MC) of a rotating black hole (BH)
with its surrounding accretion disk are discussed in detail in the
following aspects: (i) The mapping relation between the angular
coordinate on the BH horizon and the radial coordinate on the disk
is modified based on a more reasonable configuration of magnetic
field, and a condition for coexistence of the Blandford-Znajek
(BZ) and the MC process is derived. (ii) The transfer direction of
energy and angular momentum in MC process is described
equivalently by the co-rotation radius and by the flow of
electromagnetic angular momentum and redshifted energy, where the
latter is based on an assumption that the theory of BH
magnetosphere is applicable to both the BZ and MC processes. (iii)
The profile of the current on the BH horizon and that of the
current density flowing from the magnetosphere onto the horizon
are given in terms of the angular coordinate of the horizon. It is
shown that the current on the BH horizon varies with the latitude
of the horizon and is not continuous at the angular boundary
between the open and closed magnetic field lines. (iv) The MC
effects on disk radiation are discussed, and a very steep
emissivity is produced by MC process, which is consistent with the
recent \textit{XMM-Newton} observation of the nearby bright
Seyfert 1 galaxy MCG-6-30-15 by a variety of parameters of the
BH-disk system.
\end{abstract}


\label{firstpage}


\section{Introduction}

Recently magnetic coupling of a rotating black hole (BH) with its
surrounding disk has been investigated by some authors (Blandford
1999; Li 2000; Li {\&} Paczynski 2000; Li 2002a, 2002b,2002c,
hereafter Li02a, Li02b and Li02c, respectively; Wang, Xiao {\&}
Lei 2002, hereafter WXL; Wang, Lei {\&} Ma 2003, hereafter WLM),
which can be regarded as one of the variants of the
Blandford-Znajek (BZ) process proposed two decades ago (Blandford
{\&} Znajek 1977). Macdonald and Thorne (1982, hereafter MT)
reformulated and extended the BZ theory in 3+1 split of Kerr
spacetime. In MT the transportation of energy and angular momentum
from a rotating BH to a remote astrophysical load is not only
described by the flow of electromagnetic angular momentum and
energy but also by a general relativistic version of DC electronic
circuit theory based on a stationary, axisymmetric magnetosphere
anchored in the BH and its surrounding disk. In the BZ process
energy and angular momentum are extracted from a rotating BH,
which is connected with the remote astrophysical load by open
magnetic field lines. Unfortunately the remote load has been known
very few, resulting in some uncertainty in the BZ process.

With the closed magnetic field lines connecting a BH with its
surrounding disk the rotating energy of the BH provides an energy
source for disk radiation, and henceforth this energy mechanism is
referred to as magnetic coupling (MC) process. The load in MC
process is accretion disk, which is much better understood than
the remote load in the BZ process. Since the magnetic field on the
BH horizon is brought and held by the surrounding magnetized disk,
the magnetic connection between the BH and the disk is natural and
reasonable. As argued in Li02b, the magnetic connection can
produce a very steep emissivity compared to the standard
accretion, and this result is consistent with the recent
\textit{XMM-Newton} observation of the nearby bright Seyfert 1
galaxy MCG-6-30-15.

Considering the astrophysical importance of MC process, we are intended to
improve our previous model of MC process given in WXL. Based on a reasonable
consideration on the configuration of the magnetic field connecting a BH
with its surrounding disk we modify the mapping relation between the angular
coordinate on the BH horizon and the radial coordinate on the disk given in
WXL (henceforth MRWXL), and derive a condition for the coexistence of the BZ
and MC processes (henceforth CEBZMC). The transfer of energy and angular
momentum in MC process and the profile of the current on the horizon are
discussed in detail. In addition, a very steep emissivity arising from MC
process is produced in our model, being in accord with the recent
\textit{XMM-Newton} observation of the nearby bright Seyfert 1 galaxy MCG-6-30-15.

This paper is organized as follows. In Section 2 MRWXL is modified
based on more reasonable consideration of the magnetic field in BH
magnetosphere, and the angular boundary between the open and
closed field lines on the horizon is determined naturally  by the
new mapping relation and the assumption of precedence of the
magnetic flux penetrating the disk. The condition for CEBZMC is
derived and described in the 2-dimension parameter space
consisting of the BH spin and the power law index of the variation
of the magnetic field on the disk. In Section 3 the transfer
direction of energy and angular momentum between the BH and the
disk is described equivalently by the co-rotation radius based on
our recent work (WLM) and by the flow of electromagnetic angular
momentum and redshifted energy given in MT. In addition, the flux
of angular momentum and energy is discussed in detail both above
and below the equatorial plane of the BH. In Section 4 the profile
of electric current flowing on the horizon and the profile of
current density from the magnetosphere onto the horizon are
discussed based on an equivalent circuit for a unified model of
the BZ and MC processes proposed in WXL. In Section 5 the MC
effects on the disk radiation are investigated, and a very steep
emissivity is worked out to fit the recent \textit{XMM-Newton}
observation by using the new mapping relation. Finally, in Section
6, we summarize our main results.

Although load disk in MC process is much better understood than
the remote load in the BZ process, a good picture of the magnetic
connection between the BH and the disk has not been obtained. In
order to facilitate the discussion of the MC effects in an
analytic way we make the following assumptions:

\noindent (i) Assumption: The theory of a stationary, axisymmetric
magnetosphere anchored in the BH and its surrounding disk is
applicable not only to the BZ process but also to MC process. The
magnetosphere is assumed to be force-free outside the BH and the
disk.

\noindent (ii) Assumption: The disk is both stable and perfectly
conducting, and the closed magnetic field lines are frozen in the
disk. The disk is thin and Keplerian, lies in the equatorial plane
of the BH with the inner boundary being at the marginally stable
orbit.

\noindent iii) Assumption: The magnetic field is assumed to be
constant on the horizon, and to vary as a power law with the
radial coordinate of the disc.

\noindent iv) Assumption: The magnetic flux connecting a BH with
its surrounding disk takes precedence over that connecting the BH
with the remote load.

Assumption (iv) is proposed based on two reasons: (i) The magnetic
field on the horizon is brought and held by the surrounding
magnetized disk. (ii) The disk is much nearer to the BH than the
remote load.

Throughout this paper the geometric units $G = c = 1$ are used.


\section{Mapping relation and a condition for CEBZMC}

Recently a model of BH evolution was proposed in WXL by
considering CEBZMC, and the configuration of the poloidal magnetic
field is shown in Figure 1, where $r_{in} $ and $r_{out} $ are the
radii of inner and outer boundary of the MC region, respectively.
The angle $\theta _M $ indicates the angular boundary between the
open and closed field lines on the horizon, and $\theta _L $ is
the lower boundary angle for the closed field lines.


MRWXL was derived based on the conservation of magnetic flux
between the horizon and the disk with assumptions (i) - (iii).
However there are some inconsistencies with MRWXL as follows:

(i) It is impossible for one closed field line to leave and return
the horizon at the same angle $\theta _2 = \pi \mathord{\left/
{\vphantom {\pi 2}} \right. \kern-\nulldelimiterspace} 2$ on the
horizon by the symmetry of the magnetic field above and below the
equatorial plane of the BH.

(ii) The boundary condition used in MRWXL says that the normal
component of the magnetic field on the horizon is equal to that at
the inner edge of the disk, i.e. $B_\perp=B_z(r_{ms})$ . However,
it is not consistent with the estimation given by numerical
simulation: the strength of the magnetic field at the horizon is
likely greater than that in the disk (Ghosh \& Abramowicz 1997 and
the references therein).

(iii) The boundary angle $\theta _M $ should not be given at random without
any reason.

In this paper MRWXL is replaced by a new mapping relation by
considering a more reasonable configuration of the magnetic field.
Firstly, the angular $\theta _L $ is assumed less than $\pi
\mathord{\left/ {\vphantom {\pi 2}} \right.
\kern-\nulldelimiterspace} 2$ to avoid inconsistency of the closed
field line above and below the equatorial plane. Secondly, we
replace the previous boundary condition by

\begin{equation}
\label{eq1}
2\pi r_H B_ \bot = 2\pi \varpi _D \left( {r_{ms} } \right)B_z \left( {r_{ms}
} \right),
\end{equation}

\noindent where $B_ \bot $ and $B_z \left( {r_{ms} } \right)$ are
normal components of magnetic field on the horizon and at the
inner edge of the disk, respectively. In writing equation (1) we
have assumed that the two neighboring loops near the inner edge of
the disk have the same infinitesimal width, resulting in the ratio
of  $B_ \bot $ to $B_z \left( {r_{ms} } \right)$ varying from 1.8
to 3 for $0 < a_ * < 1$. The quantity $\varpi _D \left( {r_{ms} }
\right)$ is the cylindrical radius at the inner edge of the last
stable orbit radius $r_{ms} = M\chi _{ms}^2 $ and reads

\begin{equation}
\label{eq2}
\varpi _D \left( {r_{ms} } \right) = M\chi _{ms}^2 \sqrt {1 + \chi _{ms}^{ -
4} a_ * ^2 + 2\chi _{ms}^{ - 6} a_ * ^2 } ,
\end{equation}

\noindent where the parameters $M$ and $a_ * \equiv J
\mathord{\left/ {\vphantom {J {M^2}}} \right.
\kern-\nulldelimiterspace} {M^2}$ are the BH mass and spin,
respectively. Thirdly, as shown later, the boundary angle $\theta
_M $ will be determined naturally by the new mapping relation with
assumption (iv) rather than given at random.

Considering the flux tube consisting of two adjacent magnetic
surfaces ``$i$'' and ``$i + 1$'' as shown in Figure 1, we have
$\Delta \Psi_H = \Delta \Psi_D $ by continuum of magnetic flux,
i.e.,

\begin{equation}
\label{eq3}
B_ \bot 2\pi \left( {\varpi \rho } \right)_{r = r_H } d\theta = - B_z 2\pi
\left( {{\varpi \rho } \mathord{\left/ {\vphantom {{\varpi \rho } {\sqrt
\Delta }}} \right. \kern-\nulldelimiterspace} {\sqrt \Delta }}
\right)_{\theta = \pi \mathord{\left/ {\vphantom {\pi 2}} \right.
\kern-\nulldelimiterspace} 2} dr,
\end{equation}

\noindent
where we have $\varpi = \left( {\Sigma \mathord{\left/ {\vphantom {\Sigma
\rho }} \right. \kern-\nulldelimiterspace} \rho } \right)\sin \theta $. The
concerning Kerr metric parameters are given as follows (MT):

\begin{equation}
\label{eq4} \Sigma ^2 \equiv \left( {r^2 + a^2} \right)^2 -
a^2\Delta \sin ^2\theta,\quad\quad \rho ^2 \equiv r^2 + a^2\cos
^2\theta,\quad\quad \Delta \equiv r^2 + a^2 - 2Mr,
\end{equation}

\noindent So we have

\begin{equation}
\label{eq5}
\left( {\varpi \rho } \right)_{r = r_H } = \left( {\Sigma \sin \theta }
\right)_{r = r_H } = 2Mr_H \sin \theta ,
\end{equation}

\begin{equation}
\label{eq6}
\left( {{\varpi \rho } \mathord{\left/ {\vphantom {{\varpi \rho } {\sqrt
\Delta }}} \right. \kern-\nulldelimiterspace} {\sqrt \Delta }}
\right)_{\theta = \pi \mathord{\left/ {\vphantom {\pi 2}} \right.
\kern-\nulldelimiterspace} 2} = \Sigma \mathord{\left/ {\vphantom {\Sigma
{\sqrt \Delta }}} \right. \kern-\nulldelimiterspace} {\sqrt \Delta },
\end{equation}

Following Blandford (1976) we assume that $B_z $ varies as

\begin{equation}
\label{eq7}
B_z \propto \xi ^{ - n},
\end{equation}

\noindent where $\xi \equiv r \mathord{\left/ {\vphantom {r
{r_{ms} }}} \right. \kern-\nulldelimiterspace} {r_{ms} }$ is
defined as a radial parameter in terms of $r_{ms} $, and $n$ is
the power law index indicating the degree of concentration of the
magnetic field in the central region of the disk. Incorporating
equations (\ref{eq1}) and (\ref{eq7}) we have

\begin{equation}
\label{eq8}
B_z = B_ \bot \left[ {{r_H } \mathord{\left/ {\vphantom {{r_H } {\varpi _D
\left( {r_{ms} } \right)}}} \right. \kern-\nulldelimiterspace} {\varpi _D
\left( {r_{ms} } \right)}} \right]\xi ^{ - n}.
\end{equation}

Substituting equations (\ref{eq4})---(\ref{eq8}) into equation (\ref{eq3}) we have

\begin{equation}
\label{eq9}
\sin \theta d\theta = - \mbox{G}\left( {a_ * ;\xi ,n} \right)d\xi ,
\end{equation}

\noindent
where

\begin{equation}
\label{eq10} \mbox{G}\left( {a_ * ;\xi ,n} \right) =  \frac{\xi
^{1 - n}\chi _{ms}^2 \sqrt {1 + a_ * ^2 \chi _{ms}^{ - 4} \xi ^{ -
2} + 2a_ * ^2 \chi _{ms}^{ - 6} \xi ^{ - 3}} }{2\sqrt {\left( {1 +
a_ * ^2 \chi _{ms}^{ - 4} + 2a_ * ^2 \chi _{ms}^{ - 6} }
\right)\left( {1 - 2\chi _{ms}^{ - 2} \xi ^{ - 1} + a_ * ^2 \chi
_{ms}^{ - 4} \xi ^{ - 2}} \right)} }.
\end{equation}

\noindent Integrating equation (\ref{eq9}) and setting $\xi = \xi
_{in} \equiv {r_{in} } \mathord{\left/ {\vphantom {{r_{in} }
{r_{ms} }}} \right. \kern-\nulldelimiterspace} {r_{ms} } = 1$ at
$\theta _L $, we derive the new mapping relation as follows:

\begin{equation}
\label{eq11} \cos \theta - \cos \theta _L = \int_1^\xi
{\mbox{G}\left( {a_ * ;\xi ,n} \right)d\xi } ,
\end{equation}

\noindent Considering that the closed field line connects $\theta
_M $ with $\xi _{out}\equiv r_{out}/r_{ms}$, we have

\begin{equation}
\label{eq12} \cos \theta _M = \cos \theta _L + \int_1^{\xi _{out}
} {\mbox{G}\left( {a_ * ;\xi ,n} \right)} d\xi .
\end{equation}

\noindent However $\xi _{out} $ and $\theta _M $ cannot be
determined simultaneously by equation (\ref{eq12}), even if the
values of $a_* $, $n$ and $\theta _L $ are given. To determine the
boundary angle $\theta _M $, we take assumption (iv), stating that
the magnetic flux of the closed field lines takes precedence over
that due to the open field lines. Thus we have the curves of
$\xi_{out}$ and $\theta_M$ versus $a_*$ with $\theta_L=0.45\pi$
and different values of $n$ as shown in Figure 2.

From Figure 2 we obtain the following results:


(i) As shown in Figure 2a, the parameter $\theta _M $ remains zero
for $0 < a_ * < 1$, provided that the power law index $n$ is not
so great, such as $n = 3.0$. In this case the parameter $\xi
_{out} $ remains finite, varying with $a_ * $ non-monotonically
and attaining a maximum as $a_ * $ approaches unity.

(ii) As shown in Figure 2b and 2c, the parameter $\theta _M $
remains zero for the smaller value of $a_ * $, and becomes
positive for the greater value of $a_ * $, provided that $n$ is
great enough, such as $n = 4.0$\textbf{, }$4.5$\textbf{.} In this
case the parameter $\xi _{out} $ approaches infinite as soon as
$\theta _M $ becomes positive.

The above results arise from the conservation of the magnetic flux
and the precedence for the magnetic flux connecting the BH with
the disk. In our model the magnetic flux $\Psi _H $ sending out
from the BH horizon should be the sum of $\Psi _D $ and $\Psi _L
$, which are the fluxes connecting the disk and the remote load,
respectively. The maximum magnetic flux penetrating the disk is
defined by

\begin{equation}
\label{eq13}
\left( {\Psi _D } \right)_{\max } = \int_{r_{in} }^\infty {B_z 2\pi \left(
{{\varpi \rho } \mathord{\left/ {\vphantom {{\varpi \rho } {\sqrt \Delta }}}
\right. \kern-\nulldelimiterspace} {\sqrt \Delta }} \right)_{\theta = \pi
\mathord{\left/ {\vphantom {\pi 2}} \right. \kern-\nulldelimiterspace} 2}
dr} ,
\end{equation}

\noindent where the upper limit in the integral corresponds to
$\xi _{out} \to \infty $.  According to assumption (iv), we have
the two possibilities.

(1) The parameter $\xi_{out}$ is finite with $\Psi_L=0$ and
$\theta_M=0$ for $\Psi_H < (\Psi_D)_{max}$.

(2) The parameter $\xi_{out}$ is infinite with $\Psi_L \geq 0$ and
$\theta_M \geq 0$ for $\Psi_H \geq (\Psi_D)_{max}$.

It is easily to check that $\left( {\Psi _D } \right)_{\max } $ decreases
with the increasing $a_ * $ and $n$ by using equations (\ref{eq8}) and (\ref{eq13}). Thus
$\theta _M $ occurs with infinite $\xi _{out} $ as $a_ * $ and $n$ are
greater than some critical values. The existence of the positive $\theta _M
$ is just the condition for CEBZMC, which can be expressed more clearly by
the contours of $\theta _M $ in $a_ * - n$ parameter space as shown in
Figure 3.


In Figure 3 the value of $\theta _M $ is labeled beside each
contour, increasing along the direction from left-bottom to
right-top, and the thick solid line labeled zero is the critical
contour with $\theta _M = 0$. Thus the parameter space is divided
by the critical contour into two parts: (i) the right-top part for
CEBZMC, and (ii) the left-bottom part for MC process only. The
following conclusions can be obtained from Figure 3:

(i) The state of CEBZMC depends on both the BH spin $a_ * $ and
the  index $n$, and it occurs only if $a_ * $ and $n$ are great
enough to fall in the right-top part of the parameter space;

(ii) The greater is $a_ * $, the less is $n$ for the given value
of $\theta _M $. The greater are $a_ * $ and $n$, the greater is
$\theta _M $ for the given value of the lower boundary angle
$\theta _L $;

(iii) The greater is $\theta _L $, the greater is the part for
CEBZMC in the parameter space.

The requirement of CEBZMC for a greater $n$ implies that the
magnetic field is more concentrated in the central region of the
disk by equation (\ref{eq8}), although we might have $\xi _{out} $
approach infinity in this case.


\section{Transfer of energy and angular momentum}

Since the load disk in MC process is much better understood than
the remote load in the BZ process, we can obtain a better picture
of the transfer of energy and angular momentum between the BH and
the disk. In this section we are going to give two equivalent
descriptions on the transfer of energy and angular momentum in MC
process.

\subsection{Description by co-rotation radius }

In WLM the transfer direction of energy and angular momentum in MC process
is discussed in detail in terms of co-rotation radius $r_c $, and it is
defined as the radius on the disk, where the angular velocity of the disk,
$\Omega ^D$, is equal to the angular velocity of the BH, $\Omega ^H$. The
radius $r_c $ can be determined from the following equation:

\begin{equation}
\label{eq14}
{\Omega ^D} \mathord{\left/ {\vphantom {{\Omega ^D} {\Omega ^H}}} \right.
\kern-\nulldelimiterspace} {\Omega ^H} = \frac{2\left( {1 + q} \right)}{a_ *
}\left[ {\left( {\sqrt \xi \chi _{ms} } \right)^3 + a_ * } \right]^{ - 1} =
1,
\end{equation}

\noindent
and is expressed by

\begin{equation}
\label{eq15}
\xi _c \left( {a_ * } \right) \equiv {r_c } \mathord{\left/ {\vphantom {{r_c
} {r_{ms} }}} \right. \kern-\nulldelimiterspace} {r_{ms} } = \chi _{ms}^{ -
2} \left( {1 - q} \right)^{ - 1 / 3}\left( {1 + q} \right),
\end{equation}

\noindent where $q \equiv \sqrt {1 - a_ * ^2 } $ is a function of
$a_ * $. The parameter $\xi _c $ decreases monotonically with $a_
* $ as shown by the dot-dashed line in Figure 4. By using equation
(\ref{eq12}) with assumption (iv) we have the curves of $\xi
_{out} $ versus $a_ * $ for different values of the index $n$ as
shown also in Figure 4.


It is found that the MC region is divided into two parts by $\xi
_c $: the inner MC region (henceforth IMCR) for $1 < \xi < \xi _c
$ and the outer MC region (henceforth OMCR) for $\xi _c < \xi <
\xi _{out} $. Therefore energy and angular momentum are always
transferred by the closed magnetic field lines from the BH to the
disk in OMCR with $\Omega ^D < \Omega ^H$, while the transfer
direction reverses in IMCR with $\Omega ^D > \Omega ^H$. The
correlation of the BH spin with the transfer direction is given as
follows:

(i) For $0.3594 < a_ * < 1$ we have $\xi _c $ within the inner
edge, and the MC region is all in OMCR with the transfer direction
from BH to disk.

(ii) For $a_ * ^{out} < a_ * < 0.3594$ we have $1 < \xi _c < \xi
_{out} $, and the transfer is bi-directional, i.e. it is either
from BH to OMCR or from IMCR to BH.

(iii) For $0 \le a_ * < a_ * ^{out} $ we have $\xi _c > \xi _{out} $, and
the MC region is all in IMCR with the transfer direction from disk to BH.

\subsection{ Description by flow of electromagnetic angular momentum
and redshifted energy}

In MT the poloidal components of the flow of electromagnetic angular
momentum and redshifted energy in the magnetosphere are expressed as
follows:

\begin{equation}
\label{eq16}
\vec {S}_L^p = - \left( {{\vec {B} \cdot \vec {m}} \mathord{\left/
{\vphantom {{\vec {B} \cdot \vec {m}} {4\pi }}} \right.
\kern-\nulldelimiterspace} {4\pi }} \right)\vec {B}^p = \left( {I
\mathord{\left/ {\vphantom {I {2\pi \alpha c}}} \right.
\kern-\nulldelimiterspace} {2\pi \alpha c}} \right)\vec {B}^P,
\end{equation}

\begin{equation}
\label{eq17}
\vec {S}_E^P = \left( {\vec {S}_E^p } \right)_1 + \left( {\vec {S}_E^p }
\right)_2 ,
\end{equation}

\noindent
where $I$ in equation (\ref{eq16}) is the current flowing downward through an
\textbf{\textit{m}}-loop, which is related to the toroidal magnetic field
$\vec {B}^T$ by Ampere's law:

\begin{equation}
\label{eq18}
\vec {B}^T = - \frac{2I}{\alpha \varpi ^2c}\vec {m},
\end{equation}

\noindent
where $\vec {m}$ is a toroidal Killing vector as given in MT. The two terms
on RHS of equation (\ref{eq18}) are expressed by

\begin{equation}
\label{eq19}
\left( {\vec {S}_E^p } \right)_1 = \left( {{\alpha c} \mathord{\left/
{\vphantom {{\alpha c} {4\pi }}} \right. \kern-\nulldelimiterspace} {4\pi }}
\right)\left( {\vec {E}^p\times \vec {B}^T} \right),
\quad
\left( {\vec {S}_E^p } \right)_2 = \omega \vec {S}_L^p ,
\end{equation}

\noindent
where $\vec {E}^p$ in equation (\ref{eq19}) is the poloidal electric field and
satisfies

\begin{equation}
\label{eq20}
\vec {E}^p = - \left( {{\vec {v}^F} \mathord{\left/ {\vphantom {{\vec {v}^F}
c}} \right. \kern-\nulldelimiterspace} c} \right)\times \vec {B}^p
\end{equation}

In absolute-space formulation for the BZ process all the laws of physics are
expressed in terms of physical quantities measured by
``zero-angular-momentum observers'' (ZAMOs) (Bardeen, Press {\&} Teukolsky,
1972), and ZAMO angular velocity $\omega $ is expressed by

\begin{equation}
\label{eq21} \omega = 2aMr/ \Sigma ^2.
\end{equation}

In equation (\ref{eq20}) velocity $\vec {v}^F$ of the magnetic field line relative
to ZAMO is expressed as

\begin{equation}
\label{eq22}
\vec {v}^F = \alpha ^{ - 1}\left( {\Omega ^F - \omega } \right)\vec {m}
\end{equation}

Incorporating equations (\ref{eq17})---(\ref{eq22}), we have the
expression for $\vec {S}_E^P $ as follows:

\begin{equation}
\label{eq23}
\vec {S}_E^P = \Omega ^F\left( {I \mathord{\left/ {\vphantom {I {2\pi \alpha
c}}} \right. \kern-\nulldelimiterspace} {2\pi \alpha c}} \right)\vec {B}^p
\end{equation}

According to assumption (i) equations (\ref{eq16}) and
(\ref{eq23}) are also applicable to MC process. Inspecting
equations (\ref{eq16}) and (\ref{eq23}) we have the following
results.

(i) There is neither angular momentum flow nor energy flow at all unless
poloidal currents are presented $\left( {I \ne 0} \right)$.

(ii) Both angular momentum flow and energy flow along the magnetic field
lines away from the BH to the disk if the current $I$ flows downwards
through an \textbf{\textit{m}}-loop, i.e., $I > 0$, while the direction of
the two will be reversed, if the current $I$ flows upwards through an
\textbf{\textit{m}}-loop, i.e., $I < 0$.

If the description by $\vec {S}_L^p $ and $\vec {S}_E^p $ is in accord with
that by co-rotation radius $r_c $, the sign of the current $I$ should be
dependent on whether the radius $r$ is outside or inside $r_c $. We shall
discuss the sign of the current $I$ by using the equivalent circuit in a
unified model for the BZ and MC processes given in WXL as shown in Figure 5,
which is adapted to the configuration of the magnetic field in Figure 1.


In the i-loop of Figure 5 segments $PS$ (characterized by the flux $\Psi _i
)$ and $QR$ (characterized by the flux $\Psi _i + \Delta \Psi _i )$
represent two adjacent magnetic surfaces, and segments $PQ$ and $RS$
represent the BH horizon and the load sandwiched by these two
surfaces. $\Delta Z_{Hi} $ and $\Delta Z_{Li} $ are the
corresponding resistance of the BH horizon and the load,
respectively. $\Delta \varepsilon _{Hi} = \left( {{\Delta \Psi _i }
\mathord{\left/ {\vphantom {{\Delta \Psi _i } {2\pi }}} \right.
\kern-\nulldelimiterspace} {2\pi }} \right)\Omega ^H$ and $\Delta
\varepsilon _{Li} = - \left( {{\Delta \Psi _i } \mathord{\left/ {\vphantom
{{\Delta \Psi _i } {2\pi }}} \right. \kern-\nulldelimiterspace} {2\pi }}
\right)\Omega _i^F $ are electromotive forces due to the rotation of the BH
and the load, respectively. The minus sign in the expression of $\Delta
\varepsilon _{Li} $ arises from the direction of the flux. The current in
each loop is expressed as

\begin{equation}
\label{eq24}
I_i = \frac{\Delta \varepsilon _{Hi} + \Delta \varepsilon _{Li} }{\Delta
Z_{Hi} + \Delta Z_{Li} } = \left( {{\Delta \Psi _i } \mathord{\left/
{\vphantom {{\Delta \Psi _i } {2\pi }}} \right. \kern-\nulldelimiterspace}
{2\pi }} \right)\frac{\Omega ^H - \Omega _i^F }{\Delta Z_{Hi} }
\end{equation}

Suppose that an \textbf{\textit{m}}-loop is embedded the magnetic surface
QR. Inspecting Figures 1 with 5, we find that the current $I$ flowing
downwards through the \textbf{\textit{m}}-loop is simply $I_i $, i.e.,

\begin{equation}
\label{eq25} I = \left( {I_1 - I_1 } \right) + \left( {I_2 - I_2 }
\right) + \cdots + \left( {I_{i - 1} - I_{i - 1} } \right) + I_i =
I_i
\end{equation}

Suppose that the magnetic surface QR intersects with the disk at the
co-rotation radius $r_c $, we have $\Omega _i^D = \Omega ^H$ at $r_i = r_c
$. Henceforth the magnetic surface QR is referred to as co-rotation magnetic
surface (CRMS). Incorporating Figures 1 and 5 and equations (\ref{eq24}) and (\ref{eq25}),
we conclude the following results:

(i) The magnetic surface inside CRMS penetrates OMCR with
$\Omega_i^F=\Omega _i^D < \Omega ^H$ and $I > 0$. Thus $\vec
{S}_L^p $ and $\vec {S}_E^p $ are in the same direction as $\vec
{B}^P$, i.e., both angular momentum and energy are transferred
from the BH to disk.

(ii) The magnetic surface outside CRMS penetrates IMCR with
$\Omega_i^F=\Omega _i^D > \Omega ^H$ and $I < 0$. Thus $\vec
{S}_L^p $ and $\vec {S}_E^p $ are opposite to $\vec {B}^P$, i.e.,
both angular momentum and energy are transferred from the disk to
the BH.

Thus we have shown that the two descriptions for transfer of angular
momentum and energy in MC process are equivalent.

\subsection{Equivalence of the two descriptions above and below the
equatorial plane of the BH}

We can also prove that the two descriptions are equivalent both
above and below the equatorial plane of the BH by considering the
following points.

(i) Direction of $\vec {B}^p$ above and below the equatorial plane
of the BH;

(ii) Rotation of the BH with respect to the magnetic field lines
frozen in the disk;

(iii) Dragging effect of the rotating BH on the magnetic field
lines;

(iv) Equation (\ref{eq18}) relating the direction of $\vec {B}^T$
and the sign of the current $I$ through an
\textbf{\textit{m}}-loop (downwards or upwards).

We can determine the direction of $\vec {B}^T$ by points (ii) and
(iii), and determine the sign of $I$ by point (iv). Combining the
above points with the symmetry above and below the equatorial
plane, we summarize the direction of $\vec {S}_L^p $ and $\vec
{S}_E^p $ as shown in Table 1. It is shown in Table 1 the
direction of $\vec {S}_L^p $ and $\vec {S}_E^p $ depends only on
the place where the field line penetrates the disk (in OMCR or
IMCR), and the two descriptions are equivalent both above and
below the equatorial plane of the BH.

\subsection{ A further discussion on flow of energy in MC process}

From equation (\ref{eq17}) we find that $\vec {S}_E^p $ consists of two terms,
$\left( {\vec {S}_E^p } \right)_1 $ and $\left( {\vec {S}_E^p } \right)_2 $,
where the latter has the same direction as $\vec {S}_L^p $, and the former
is related to the directions of $\vec {E}^p$ and $\vec {B}^T$. We are going
to give a discussion on the direction of $\left( {\vec {S}_E^p } \right)_1 $
in detail.

\textbf{Direction of }$\vec {E}^p$\textbf{:} From equation (\ref{eq20}) we know that
the direction of $\vec {E}^p$ is normal to $\vec {B}^p$, depending on the
cross product of $\vec {v}^F$ and $\vec {B}^p$. While $\vec {v}^F$ depends
on the difference between $\Omega ^F$ and ZAMO angular velocity $\omega $ by
equation (\ref{eq22}). Since the closed magnetic field lines are frozen in the disk,
so $\Omega ^F$ depends on the place where the field line penetrates the
disk, i.e.,

\begin{equation}
\label{eq26} \Omega ^F = \Omega ^D = \frac{1}{M\left( {\xi ^{3
\mathord{\left/ {\vphantom {3 2}} \right.
\kern-\nulldelimiterspace} 2}\chi _{ms}^3 + a_ * } \right)}.
\end{equation}

From equation (\ref{eq21}) we know that $\omega $ depends on ZAMO
coordinates $\left( {r,\mbox{ }\theta } \right)$ in BH
magnetosphere, and it satisfies $\Omega ^H \ge \omega $, where the
equality holds as ZAMOs reach the horizon. Generally speaking, the
farther are ZAMOs away from the horizon, the less is $\omega $.
Incorporating equations (\ref{eq20}), (\ref{eq21}), (\ref{eq22})
and (\ref{eq26}), we find that the direction of $\vec {E}^p$ is a
little complicated, depending on the position of ZAMOs at the
field line and on the position of the foot of the field line on
the disk.

\textbf{Direction of }$\vec {B}^T$\textbf{: }The direction of $\vec {B}^T$
has been shown in Table 1. Considering the dragging effect of the rotating
BH on the magnetic field, the direction of $\vec {B}^T$ depends on the
difference between $\Omega ^H$ and $\Omega ^F$. The toroidal magnetic field
$\vec {B}^T$ is in the same direction as vector $\vec {m}$ for $\Omega ^H <
\Omega ^F$, and it is opposite to $\vec {m}$ for $\Omega ^H > \Omega ^F$.

Based on the above discussion we can determine the direction of
$\left( {\vec {S}_E^p } \right)_1 $ and those of the concerning
quantities according to the three possibilities for the sequence
in magnitude of $\Omega ^H$, $\omega $ and $\Omega ^F$ as listed
in Table 2. It is found from Table 1 and Table 2 that $\left(
{\vec {S}_E^p } \right)_1 $ is in the same direction as $\vec
{S}_E^p $ in cases A, B, D and E, while it is opposite to $\vec
{S}_E^p $ in cases C and F. Considering that both $\vec {S}_E^p $
and $\left( {\vec {S}_E^p } \right)_2 $ are in the same direction,
we infer that $\left( {\vec {S}_E^p } \right)_1 $ is dominated by
$\left( {\vec {S}_E^p } \right)_2 $ in cases C and F.

\section{Profile of electric current in BH magnetosphere}

Based on MT and our model depicted in Figures 1 and 5 we can discuss the
profile of the current on the BH horizon and that of the current density
flowing onto the horizon from BH magnetosphere. As argued in MT the region
outside the disk and the BH can be regarded as a force-free region, and the
following equation is satisfied:

\begin{equation}
\label{eq27}
\rho _e \vec {E} + \left( {\vec {j} \mathord{\left/ {\vphantom {\vec {j} c}}
\right. \kern-\nulldelimiterspace} c} \right)\times \vec {B} = 0
\end{equation}

According to Assumption (i) equation (\ref{eq27}) is applicable to
MC process, from which we infer that the poloidal component of the
current density $\vec {j}^p$ is parallel to $\vec {B}^p$ by
considering $\vec {E}^T = 0$ in the stationary magnetosphere. We
can derive the profile of the current on the horizon and the
profile of the current density $j_{MH} $ in terms of the latitude
of the horizon.

The equivalent circuit in Figure 5 is applicable to both the BZ and MC
processes, where the current in each loop circuit is exactly the current on
the horizon. Substituting $\Delta \Psi = B_H 2\pi \varpi \Delta l$, $\Delta
Z_H = {2\rho \Delta \theta } \mathord{\left/ {\vphantom {{2\rho \Delta
\theta } \varpi }} \right. \kern-\nulldelimiterspace} \varpi $ and the
concerning parameters in the Kerr metric into equation (\ref{eq24}), we have the
current on the horizon due to the BZ and MC processes as follows:

\begin{equation}
\label{eq28}
I_{HBZ} = B_H M\frac{a_ * \left( {1 - k} \right)}{2\csc ^2\theta - \left( {1
- q} \right)} = I_0 \frac{a_ * \left( {1 - k} \right)}{2\csc ^2\theta -
\left( {1 - q} \right)},
\end{equation}

\begin{equation}
\label{eq29}
I_{HMC} = B_H M\frac{a_ * \left( {1 - \beta } \right)}{2\csc ^2\theta -
\left( {1 - q} \right)} = I_0 \frac{a_ * \left( {1 - \beta } \right)}{2\csc
^2\theta - \left( {1 - q} \right)},
\end{equation}

\noindent where $I_0 = B_H \left( {{GM} \mathord{\left/ {\vphantom
{{GM} c}} \right. \kern-\nulldelimiterspace} c} \right) \approx
1.48\times 10^{10}B_4 \left( {M \mathord{\left/ {\vphantom {M {M_
\odot }}} \right. \kern-\nulldelimiterspace} {M_ \odot }}
\right)A$, and $B_4 $ is the strength of the magnetic field on the
horizon in the unit of $10^4 gauss$. The two parameters $k \equiv
{\Omega ^F} \mathord{\left/ {\vphantom {{\Omega ^F} {\Omega ^H}}}
\right. \kern-\nulldelimiterspace} {\Omega ^H}$ and $\beta \equiv
{\Omega ^F} \mathord{\left/ {\vphantom {{\Omega ^F} {\Omega ^H}}}
\right. \kern-\nulldelimiterspace} {\Omega ^H} = {\Omega ^D}
\mathord{\left/ {\vphantom {{\Omega ^D} {\Omega ^H}}} \right.
\kern-\nulldelimiterspace} {\Omega ^H}$ are the ratios of the
angular velocity of the field lines to that of the BH in the BZ
and MC processes, respectively. The value range of $k$ is $0 < k <
1$, which cannot be determined precisely due to lack of knowledge
of the remote load. Although people know very few about the remote
load in the BZ process, the ratio $k$ is assumed to be 0.5 for the
optimal BZ power (MT). Compared with the BZ process the value of
$\beta $ can be determined precisely in terms of $a_ * $ and $\xi
$ by the following equation:

\begin{equation}
\label{eq30} \beta \equiv {\Omega _D } \mathord{\left/ {\vphantom
{{\Omega _D } {\Omega _H }}} \right. \kern-\nulldelimiterspace}
{\Omega _H } = \frac{2\left( {1 + q} \right)}{a_ * }\left[ {\left(
{\sqrt \xi \chi _{ms} } \right)^3 + a_ * } \right]^{ - 1}.
\end{equation}

Incorporating the new mapping relation (\ref{eq11}) and equations
(\ref{eq29}) and (\ref{eq30}) we have the curves of ${I_{HMC} }
\mathord{\left/ {\vphantom {{I_{HMC} } {I_0 }}} \right.
\kern-\nulldelimiterspace} {I_0 }$ versus $\theta $ from $\theta
_M $ to $\theta _L = 0.45\pi $ for the given values of the power
law index $n$ as shown in Figure 6.


From Figure 6 we find the following results:

(i) The current $I_{HMC} $ varies with $\theta $ on the horizon, where
${I_{HMC} } \mathord{\left/ {\vphantom {{I_{HMC} } {I_0 }}} \right.
\kern-\nulldelimiterspace} {I_0 } > 0$ indicates the current flowing from
high to low latitude, and ${I_{HMC} } \mathord{\left/ {\vphantom {{I_{HMC} }
{I_0 }}} \right. \kern-\nulldelimiterspace} {I_0 } < 0$ indicates the
current flowing from low to high latitude;

(ii) From equation (\ref{eq29}) we know that the sign of ${I_{HMC} } \mathord{\left/
{\vphantom {{I_{HMC} } {I_0 }}} \right. \kern-\nulldelimiterspace} {I_0 }$
depends on the MC region with which its angular region is connected. As
argued in Subsection 3.1, we have ${I_{HMC} } \mathord{\left/ {\vphantom
{{I_{HMC} } {I_0 }}} \right. \kern-\nulldelimiterspace} {I_0 } > 0$ for $a_
* ^c < a_ * < 1$ with the angular region corresponding to OMCR as shown in
Figure 6a, and have ${I_{HMC} } \mathord{\left/ {\vphantom {{I_{HMC} } {I_0
}}} \right. \kern-\nulldelimiterspace} {I_0 } < 0$ for $0 < a_ * < a_ * ^c $
with the angular region corresponding to IMCR as shown in Figure 6c. Both
${I_{HMC} } \mathord{\left/ {\vphantom {{I_{HMC} } {I_0 }}} \right.
\kern-\nulldelimiterspace} {I_0 } > 0$ and ${I_{HMC} } \mathord{\left/
{\vphantom {{I_{HMC} } {I_0 }}} \right. \kern-\nulldelimiterspace} {I_0 } <
0$ occur for $a_ * ^c < a_ * < 0.3594$, since both OMCR and IMCR exist in
this case as shown in Figure 6b.

From the above discussion, the current ${I_{HBZ} } \mathord{\left/
{\vphantom {{I_{HBZ} } {I_0 }}} \right. \kern-\nulldelimiterspace} {I_0 }$
exists only in CEBZMC with $\theta _M > 0$. An interesting issue is whether
the current is continuous at the boundary angle $\theta _M $ on the horizon.
Based on the condition for CEBZMC we know that $\theta _M $ corresponds to
$\xi _{out} \to \infty $, so we have $\beta \to 0$ at $\theta _M $ by
equation (\ref{eq30}) except the Schwarzschild BH with $a_ * = 0$. Thus the
difference between ${I_{HMC} } \mathord{\left/ {\vphantom {{I_{HMC} } {I_0
}}} \right. \kern-\nulldelimiterspace} {I_0 }$ and ${I_{HBZ} }
\mathord{\left/ {\vphantom {{I_{HBZ} } {I_0 }}} \right.
\kern-\nulldelimiterspace} {I_0 }$ is given by

\begin{equation}
\label{eq31} {I_{\theta M} } \mathord{\left/ {\vphantom
{{I_{\theta M} } {I_0 }}} \right. \kern-\nulldelimiterspace} {I_0
} = \frac{a_ * k}{2\csc ^2\theta _M - \left( {1 - q} \right)}.
\end{equation}

It is found from equation (\ref{eq31}) that $I_{\theta M}
> 0$ will hold for CEBZMC with $\theta _M > 0$, and the current
$I_{\theta M} $ flows from the magnetosphere onto the BH horizon,
which could be regarded as a beam of electrons emitting from the
horizon to the magnetosphere. Inspecting equations (\ref{eq28})
and (\ref{eq29}), we find that the current $I_{\theta M} $ arises
from the difference between the two parameters $k$ and $\beta $ at
$\theta _M $, which correspond to the two different kinds of
loads.

Still by the conservation of current we can calculate the current
$I_{MH} $ flowing from the magnetosphere onto the horizon in the
range $\theta _M < \theta < \theta _L $, since $I_{HMC} $ is not
constant on the horizon. Inspecting Figure 5, we know that $I_{MH}
$ arises from the difference between the current of the two
adjacent loops. Incorporating the new mapping relation
(\ref{eq11}) and equations (\ref{eq29}) and (\ref{eq30}), we have
the current density $j_{MH} $ flowing from the magnetosphere onto
the horizon as follows:

\begin{equation}
\label{eq32}
\begin{array}{l}
 j_{MH} = \frac{1}{2\pi \left( {\varpi \rho } \right)_{r = r_H }
}\frac{dI}{d\theta } \\ \\\quad\quad = j_0 \frac{M\Omega _H }{2 -
\sin ^2\theta \left( {1 - q} \right)}\left[ {\frac{4\left( {1 -
\beta } \right)\cos \theta }{2 - \sin ^2\theta \left( {1
- q} \right)} - \sin \theta \frac{d\beta }{d\theta }} \right] \\
 \end{array}
\end{equation}

\noindent
where $j_0 \equiv \frac{B_H c^3}{2\pi GM} = 0.108\times B_4 \left( {M
\mathord{\left/ {\vphantom {M {M_ \odot }}} \right.
\kern-\nulldelimiterspace} {M_ \odot }} \right)^{ - 1}A \cdot cm^{ - 2}$,
and

\begin{equation}
\label{eq33}
\frac{d\beta }{d\theta } = \frac{3\left( {1 + q} \right)\chi _{ms}^3 \xi ^{1
\mathord{\left/ {\vphantom {1 2}} \right. \kern-\nulldelimiterspace} 2}}{a_
* \left( {\xi ^{3 \mathord{\left/ {\vphantom {3 2}} \right.
\kern-\nulldelimiterspace} 2}\chi _{ms}^3 + a_ * } \right)^2}\frac{\sin
\theta }{\mbox{G}\left( {a_ * ;\xi ,n} \right)}
\end{equation}

By using equations (\ref{eq32}) and (\ref{eq33}) we have the curve of ${j_{MH} }
\mathord{\left/ {\vphantom {{j_{MH} } {j_0 }}} \right.
\kern-\nulldelimiterspace} {j_0 }$ versus $\theta $ as shown in Figure 7.


Comparing Figure 6 with Figure 7, we find that the positive value of
${j_{MH} } \mathord{\left/ {\vphantom {{j_{MH} } {j_0 }}} \right.
\kern-\nulldelimiterspace} {j_0 }$ always accompanies with the increasing
${I_{HMC} } \mathord{\left/ {\vphantom {{I_{HMC} } {I_0 }}} \right.
\kern-\nulldelimiterspace} {I_0 }$ for the given values of $a_ * $ and $n$,
while the negative value of ${j_{MH} } \mathord{\left/ {\vphantom {{j_{MH} }
{j_0 }}} \right. \kern-\nulldelimiterspace} {j_0 }$ always occurs
simultaneously with the decreasing ${I_{HMC} } \mathord{\left/ {\vphantom
{{I_{HMC} } {I_0 }}} \right. \kern-\nulldelimiterspace} {I_0 }$. These
results are just required by the conservation of the current in BH
magnetosphere.

\section{MC process and disk radiation}

Based on the three conservation laws of mass, energy and angular momentum,
Li derived the radiation flux due to the MC effects as follows (Li02a):

\begin{equation}
\label{eq34}
F_{MC} = - \frac{d\Omega _D }{rdr}\left( {E^ + - \Omega _D L^ + } \right)^{
- 2}\int_{r_{ms} }^r {\left( {E^ + - \Omega _D L^ + } \right)Hrdr} ,
\end{equation}

\noindent
where $H$ is the angular momentum flux transferred between the BH and the
disk, and is assumed to be distributed from $r = r_{ms} $ to $r = r_{out} $
with a power law:

\begin{equation}
\label{eq35} H = \left\{ {\begin{array}{l}
 Ar^m,\mbox{ }r_{ms} < r < r_{out} \\
 0,\quad\quad\quad\mbox{ }r > r_{out} \\
 \end{array}} \right.
\end{equation}

\noindent
where $A$ is regarded as a constant in Li02b. It is shown in Li02b that the
magnetic coupling between a black hole and a disk can produce a very steep
emissivity with index $\alpha = 4.3 \sim 5.0$, where the emissivity index is
defined as

\begin{equation}
\label{eq36}
\alpha \equiv - {d\ln F} \mathord{\left/ {\vphantom {{d\ln F} {d\ln r}}}
\right. \kern-\nulldelimiterspace} {d\ln r}.
\end{equation}

This result is consistent with the recent \textit{XMM-Newton} observation of the nearby bright
Seyfert 1 galaxy MCG-6-30-15, being regarded as one of the observational
signatures of the magnetic coupling between a rotating BH and the
surrounding disk.

Unfortunately, the derivation of equation (\ref{eq35}) was not
given in Li02a and Li02b, which should be related to the torque
exerted on the BH in MC process. In WXL the torque exerted on the
BH is expressed by

\begin{equation}
\label{eq37}
 T_{MC}/T_0=4a_*(1+q)\int_{\theta_M}^{\theta_L}\frac{(1-\beta)sin^3\theta d\theta}{2-(1-q)sin^2\theta},
\end{equation}

\noindent
where

\begin{equation}
\label{eq38}
T_0 = B_4^2 M_8^3 \times 3.26\times 10^{47}g \cdot cm^2 \cdot s^{ - 2},
\end{equation}

\noindent
and $M_8 $ is the BH mass in the unit of $10^8M_ \odot $. The flux $H$ can
be derived from $T_{MC} $ by the conservation of angular momentum as
follows:

\begin{equation}
\label{eq39}
{\partial T_{MC} } \mathord{\left/ {\vphantom {{\partial T_{MC} } {\partial
r}}} \right. \kern-\nulldelimiterspace} {\partial r} = 4\pi rH = -
\frac{4T_0 a_ * \left( {1 + q} \right)\left( {1 - \beta } \right)\sin
^3\theta }{2 - \left( {1 - q} \right)\sin ^2\theta }\frac{\partial \theta
}{\partial r},
\end{equation}

\noindent
where ${\partial \theta } \mathord{\left/ {\vphantom {{\partial \theta }
{\partial r}}} \right. \kern-\nulldelimiterspace} {\partial r}$ is related
to the new mapping relation (\ref{eq11}) by

\begin{equation}
\label{eq40}
{\partial \theta } \mathord{\left/ {\vphantom {{\partial \theta } {\partial
r}}} \right. \kern-\nulldelimiterspace} {\partial r} = \left( {{\partial
\theta } \mathord{\left/ {\vphantom {{\partial \theta } {\partial \xi }}}
\right. \kern-\nulldelimiterspace} {\partial \xi }} \right)\left( {{\partial
\xi } \mathord{\left/ {\vphantom {{\partial \xi } {\partial r}}} \right.
\kern-\nulldelimiterspace} {\partial r}} \right) = - \frac{\mbox{G}\left(
{a_ * ;\xi ,n} \right)}{r_{ms} \sin \theta }.
\end{equation}

\noindent Incorporating equations (\ref{eq39}) and (\ref{eq40}),
we have

\begin{equation}
\label{eq41}
{H\left( {a_ * ;\xi ,n} \right)} \mathord{\left/ {\vphantom {{H\left( {a_ *
;\xi ,n} \right)} {H_0 }}} \right. \kern-\nulldelimiterspace} {H_0 } =
\left\{ {\begin{array}{l}
 A\left( {a_ * ,\xi } \right)\xi ^{ - n}, 1 < \xi  < \xi
_{out} \\
 \quad 0,\quad\quad\quad\quad\quad \xi > \xi _{out}  \\
 \end{array}} \right.
\end{equation}

\noindent
where $H_0 = 1.48\times 10^{21}\times B_4^2 M_8 \mbox{ }g \cdot s^{ - 2}$,
and

\begin{equation}
\label{eq42}
\left\{ {\begin{array}{l}
 A\left( {a_ * ,\xi } \right) = \frac{a_ * \left( {1 - \beta } \right)\left(
{1 + q} \right)}{2\pi \chi _{ms}^2 \left[ {2\csc ^2\theta - \left(
{1 - q} \right)} \right]}F_A \left( {a_ * ,\xi } \right) \\ \\
 F_A \left( {a_ * ,\xi } \right) = \frac{\sqrt {1 + a_ * ^2 \chi _{ms}^{ -
4} \xi ^{ - 2} + 2a_ * ^2 \chi _{ms}^{ - 6} \xi ^{ - 3}} }{\sqrt {\left( {1
+ a_ * ^2 \chi _{ms}^{ - 4} + 2a_ * ^2 \chi _{ms}^{ - 6} } \right)\left( {1
- 2\chi _{ms}^{ - 2} \xi ^{ - 1} + a_ * ^2 \chi _{ms}^{ - 4} \xi ^{ - 2}}
\right)} } \\
 \end{array}} \right.
\end{equation}

Thus we derive the expression (\ref{eq41}) for the function $H$,
where $m = - n$, and the coefficient $A\left( {a_ * ,\xi }
\right)$ is dependent on $a_ * $ and $\xi $ rather than a constant
given in equation (\ref{eq35}).

Incorporating equations (\ref{eq34}), (\ref{eq36}) and
(\ref{eq41}), we have the curves of the emissivity index $\alpha $
versus $\lg \left( {r \mathord{\left/ {\vphantom {r M}} \right.
\kern-\nulldelimiterspace} M} \right)$ for the different values of
$\theta _L $ and $n$ in Figure 8. As shown in the shaded region of
Figure 8, we find that the recent \textit{XMM-Newton} observation
of the nearby bright Seyfert 1 galaxy MCG-6-30-15 can be simulated
by the new mapping relation (\ref{eq11}) with three parameters:
(i) the BH spin $a_ * $, (ii) the power law index $n$, and (iii)
the lower boundary angle $\theta _L $. While the index produced by
a standard accretion disc (SAD) is far below the shaded region as
shown by the solid lines in Figure 8. More parameters suitable to
the observation are listed in Table 3, from which we find that the
condition for CEBZMC is well satisfied by these values of the
parameters $a_ * $ and $n$. According to the condition for CEBZMC
derived in our model we expect that an jet might be produced
nearby bright Seyfert 1 galaxy MCG-6-30-15 by the BZ process
accompanying with MC process.


\section{Summary}

In this paper the MC effects are discussed based on a stationary,
axisymmetric magnetosphere anchored in the BH and its surrounding
disk. Since we know very few about the magnetic field connecting
BH with disk, the discussion is based on some simplified
assumptions. It is known that electromagnetic field is related to
electric charge density and current density in  BH magnetosphere
by a 'stream equation' containing the 'stream function' proposed
in MT.  Unfortunately it is a rather complicated task to solve
analytically the `stream equation' with the boundary conditions at
horizon and disk surface.

Recently the magnetic connection of a Kerr BH with disk and the
resulting transportation of energy and angular momentum are
discussed analytically based on a toy model in Li02c, where the
poloidal magnetic field connecting the BH with the disk is
naturally produced by a single toroidal current flowing around the
BH in the equatorial plane. Although the existence of the single
toroidal current needs further explanation, it is a step towards
the goal of the origin of the magnetic connection. Although the
configuration of the magnetic field in Li02c is different from
that described in our model, the very steep emissivities in the
disk are produced in both models. So we think that the very steep
emissivities can be regarded as the main feature of the magnetic
connection between the BH and the disk.

The main results of our model are summarized as follows. (i) A
consistent picture about the configuration of the magnetic field
is depicted by Figure 1 with the equivalent circuit in Figure 5,
and the transfer of energy and angular momentum along the closed
field lines in MC process is described clearly and virtually in
the two equivalent descriptions. (ii) The condition for CEBZMC is
derived naturally based on some reasonable assumptions, which
might be helpful in explaining the high energy radiation from
BH-disk systems by combining the two mechanisms. (iii) The profile
of the current on the horizon is calculated, and its continuity at
the boundary angle   is discussed based on the equivalent circuit.
(iv) The MC effects on the disk radiation and the emissivity index
are investigated, and the results turn out to be in accord with
the observation.

In our model the most important parameters of the system are the
BH spin $a_*$  and the power law index $n$. The importance of BH
spin $a_*$ rests in the two aspects: (i) The ratio of the rotating
energy to the total energy of a Kerr BH is a function of $a_*$,
and it increases monotonically with the increasing $a_*$ (Wald
1984). (ii) The extracting powers in the BZ and MC processes are
all proportional to $a_*^2$ (WXL). The power law index $n$ plays
an important role in adjusting the profile of the magnetic field
in the two aspects: (i) The more is the value of $n$, the more is
the magnetic field concentrated in the central region of the disk;
(ii) The more is the value of $n$, the less is the maximum
magnetic flux penetrating the disk, and the more is the boundary
angle $\theta_M $, and the more is the BZ power due to the effects
of the open field lines on the horizon.

In this paper, we fail to discuss the MC effects on disk accretion
rate by the following consideration: The variation of disk
accretion rate due to the transfer of angular momentum is probably
a dynamical process rather than a stationary one, which is much
more complicated than the case we discuss in this paper. We shall
discuss the MC effects on accretion rate in the future work.

\acknowledgments
 {\bf Acknowledgements:}
 This work is supported by the National Natural
Science Foundation of China under Grant Numbers 10173004 and
10121503.


\begin{figure}

\epsscale{0.5}
\begin{center}
\plotone{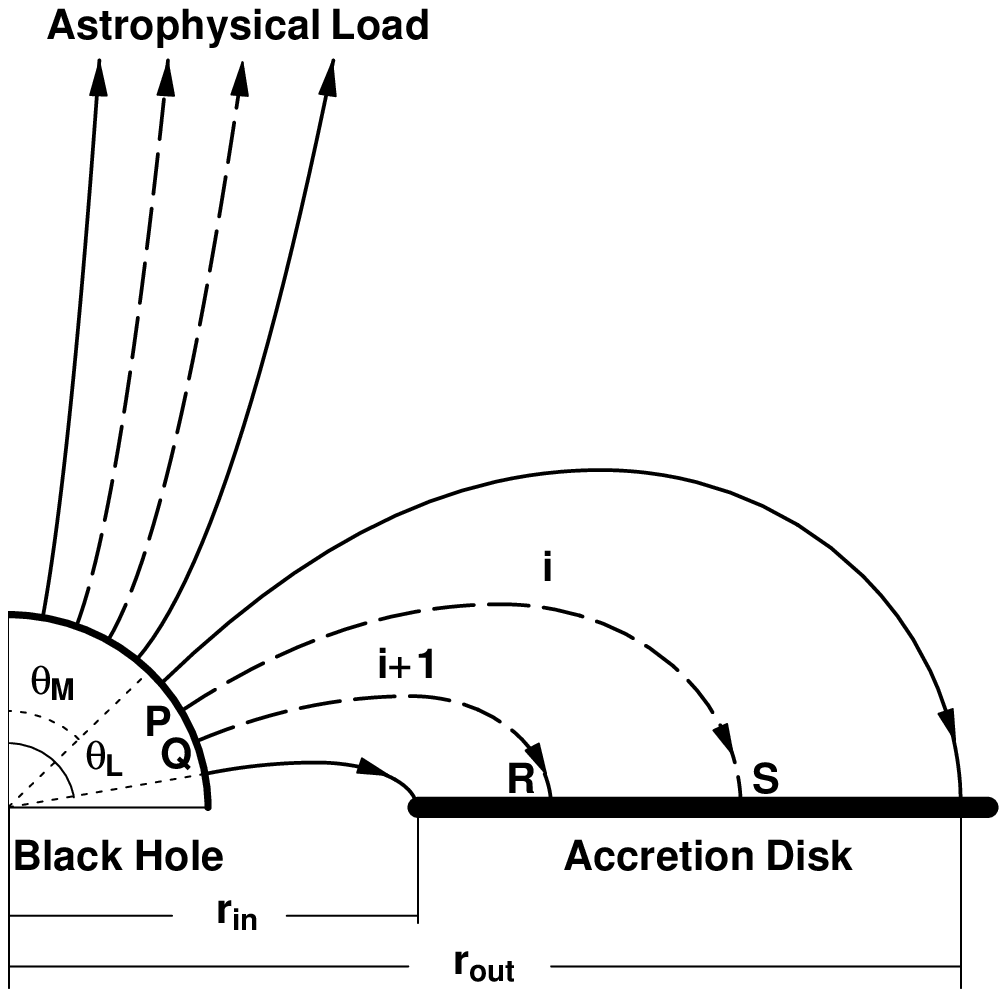}
\end{center}

\caption{The poloidal magnetic field connecting a rotating BH with
remote astrophysical load and a surrounding disk} \label{fig1}

\end{figure}


\begin{center}
\begin{figure}
\epsscale{0.35}
\begin{center}
 \plotone{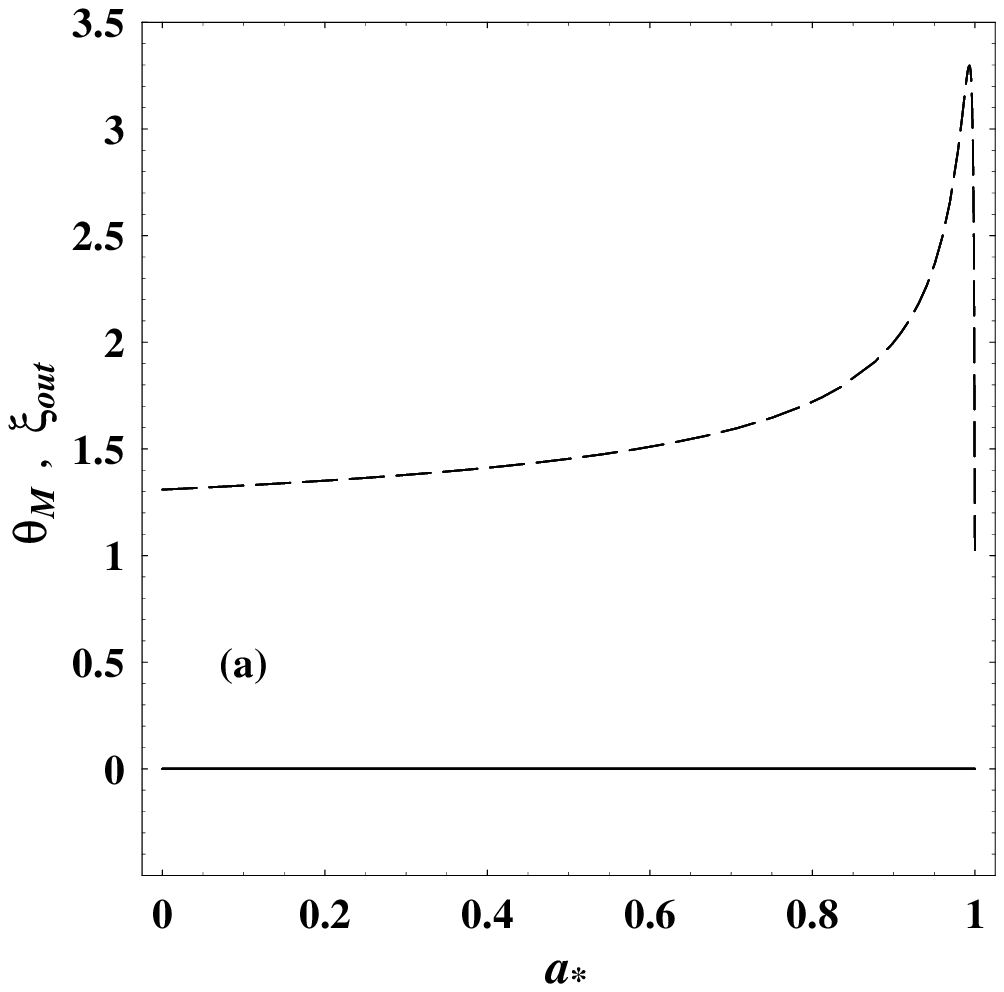}
\end{center}
\begin{center}
 \epsscale{0.35}
  \plotone{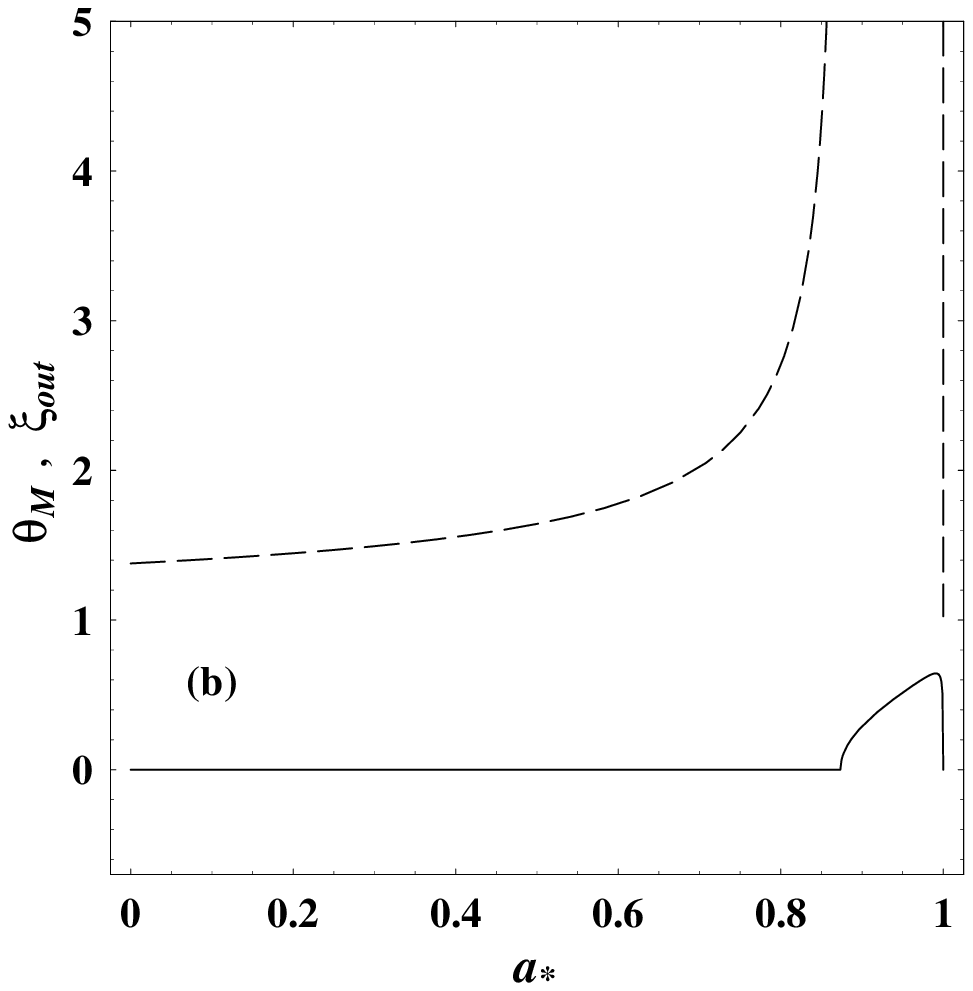}
\end{center}
\begin{center}
 \epsscale{0.35}
 \plotone{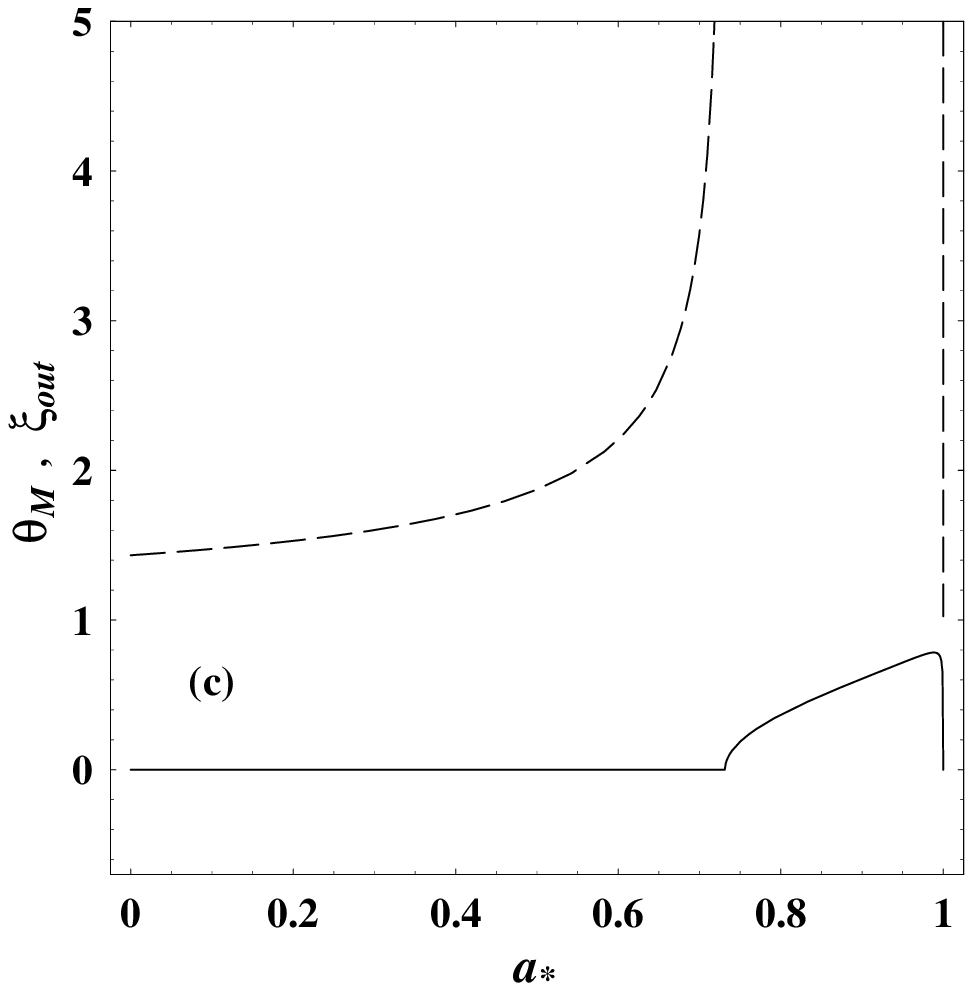}
 \end{center}
\caption{The curves of $\theta _M $ (solid line) and $\xi _{out} $
(dashed line) versus $a_ * $ with $\theta _L = 0.45\pi $ and
different values of $n$: (a) $n = 3.0$, (b) $n = 4.0$, (c) $n =
4.5$.} \label{fig2}
\end{figure}
\end{center}


\begin{figure}
\begin{center}
\epsscale{0.5}
 \plotone{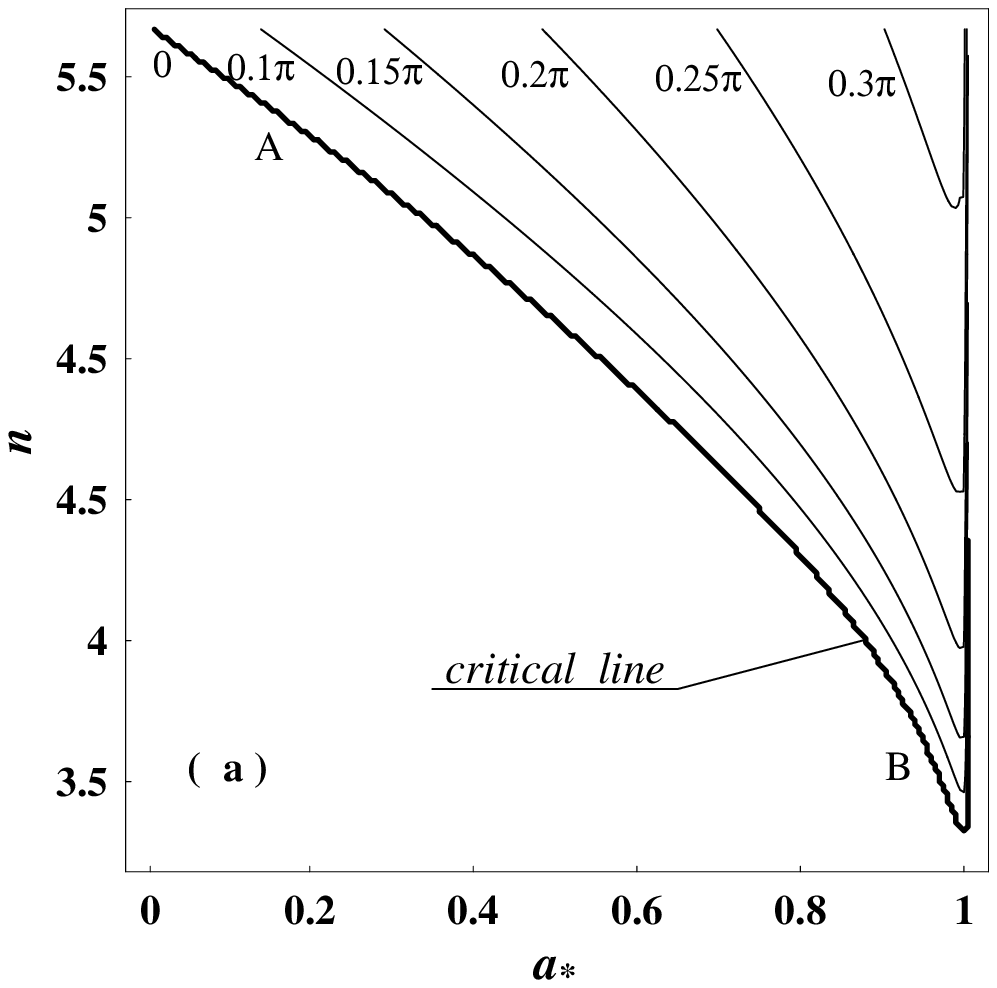}
 \end{center}
 \begin{center}
\epsscale{0.5}
 \plotone{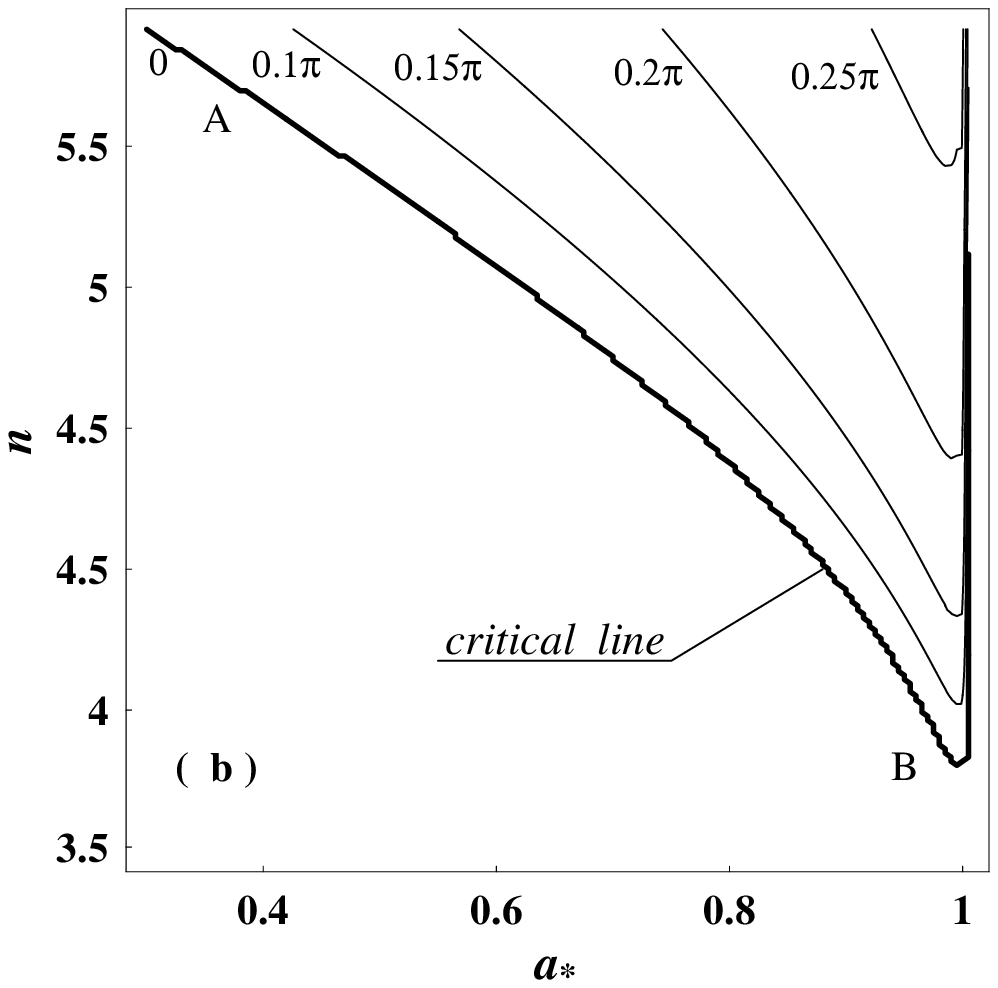}
 \caption{The contours of constant values of $\theta _M $ in $a_ * - n$
parameter space with (a) $\theta _L = 0.45\pi $, (b) $\theta _L =
0.40\pi $.}
 \label{fig3}
\end{center}
\end{figure}


\begin{figure}
\begin{center}
\epsscale{0.4} \plotone{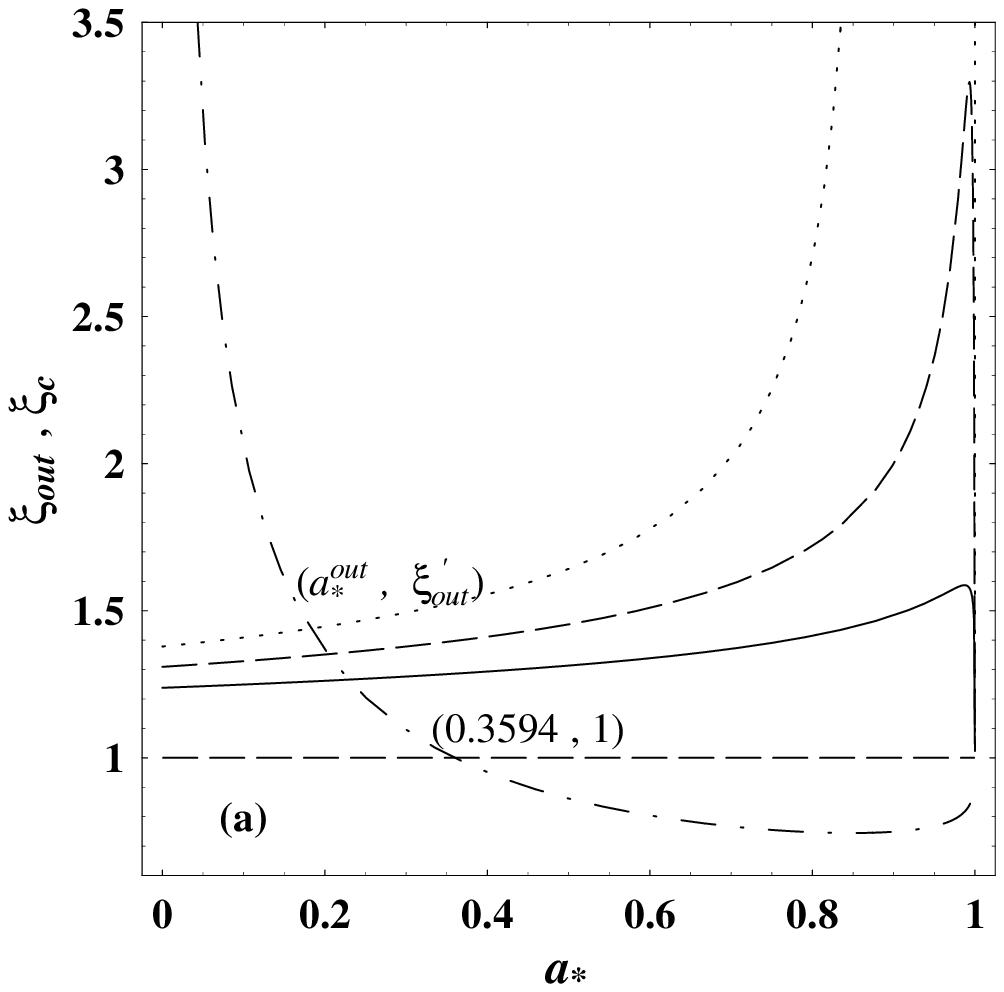}
\end{center}
\begin{center}
\epsscale{0.4} \plotone{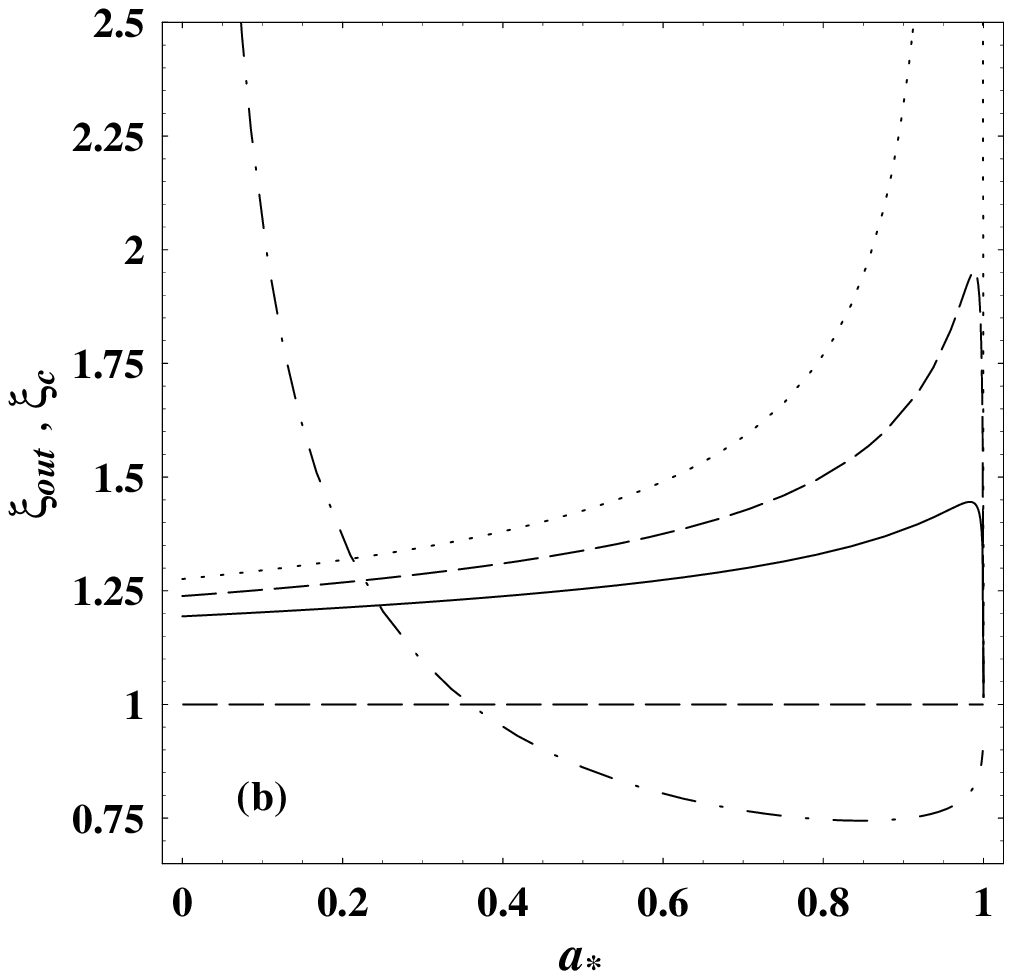}
\end{center}
\caption{The curves of $\xi _c $ (dot-dashed line) and $\xi _{out}
$ versus $a_ * $ for $0 < a_ * < 1$ with $n = 1.1,\mbox{ 3.0}$ and
$4.0$ in solid, dashed and dotted lines, respectively, (a) $\theta
_L = 0.45\pi $, (b) $\theta _L = 0.40\pi $.} \label{fig4}
\end{figure}

\clearpage
\begin{figure}
\epsscale{0.8}
\begin{center}
\plotone{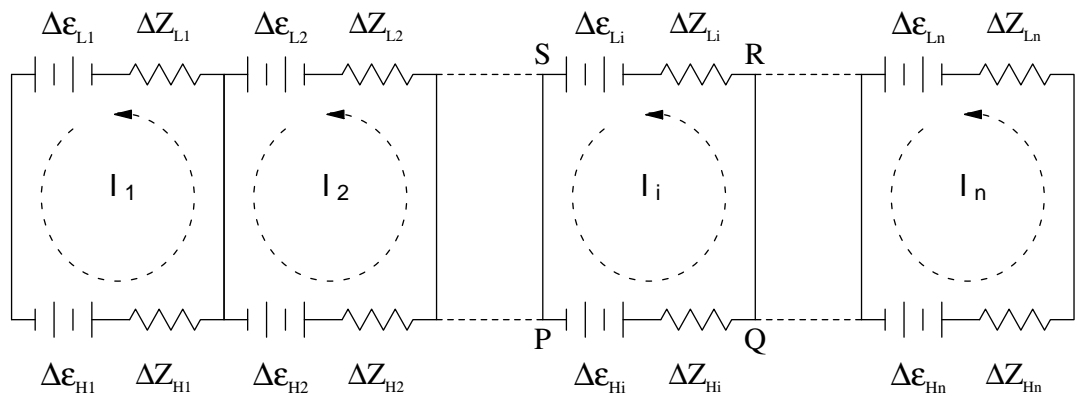}
 \label{fig5}
\end{center}
\caption{An equivalent circuit for a unified model for the BZ and
MC processes}
\end{figure}

\clearpage
\begin{figure}
\epsscale{0.35}
\begin{center}
\plotone{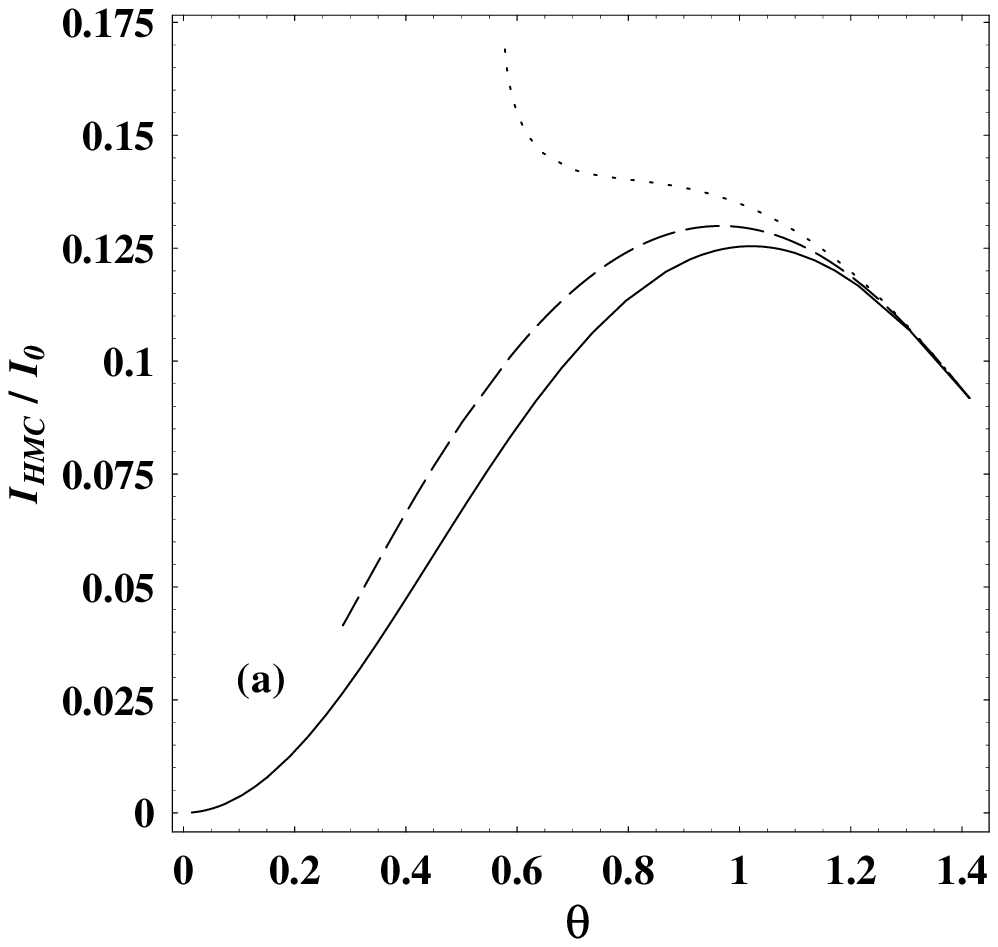}
\end{center}
\begin{center}
\plotone{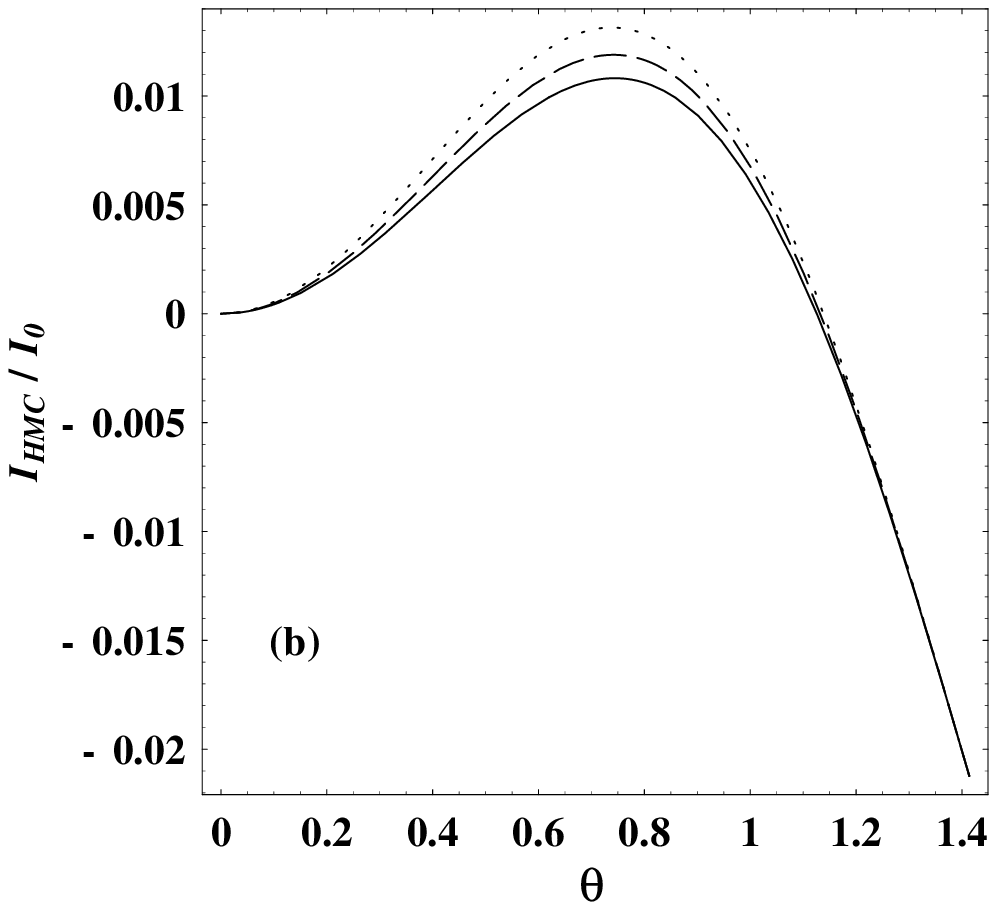}
\end{center}
\begin{center}
\plotone{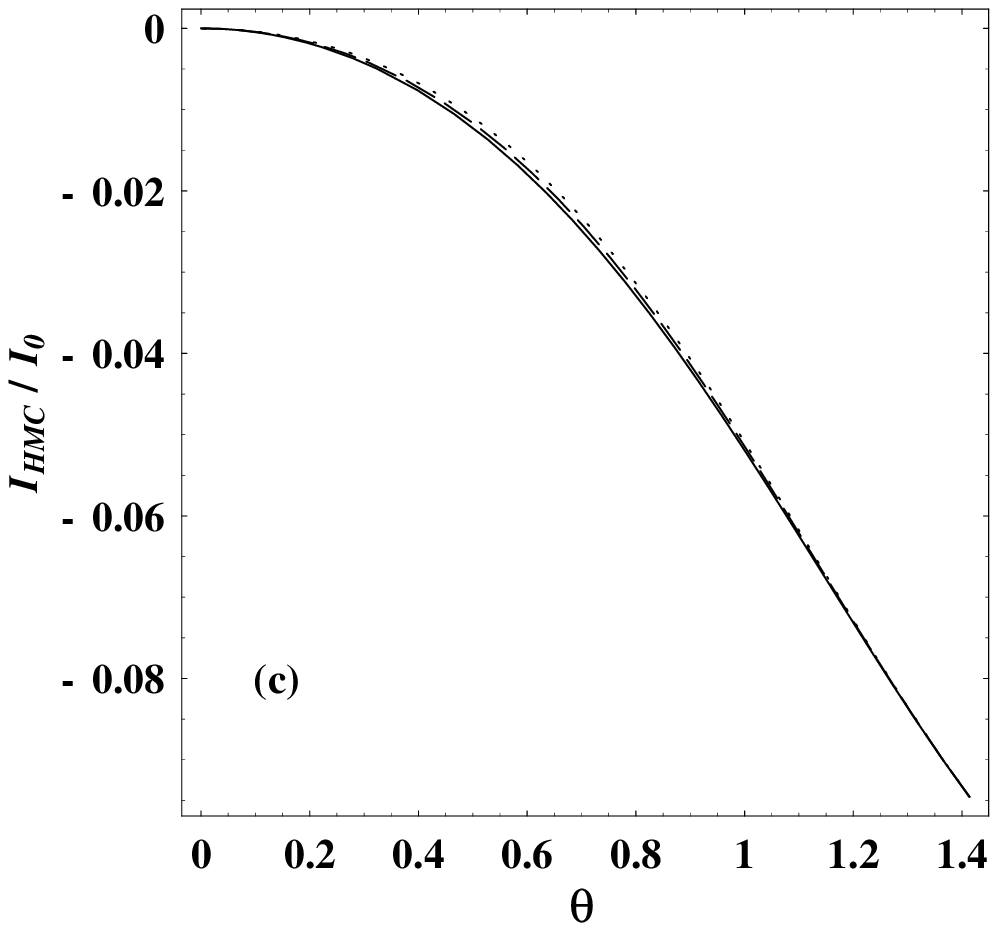}
\end{center}
\caption{The curves of the current $I_{HMC}/I_0 $ versus $\theta $
from $\theta _M $ to $\theta _L = 0.45\pi $ with $n = 3.0,\mbox{
4.0}$ and $4.5$ in solid, dashed and dotted lines, respectively.}
\label{fig6}
\end{figure}

\clearpage
\begin{figure}
\epsscale{0.35}
\begin{center}
\plotone{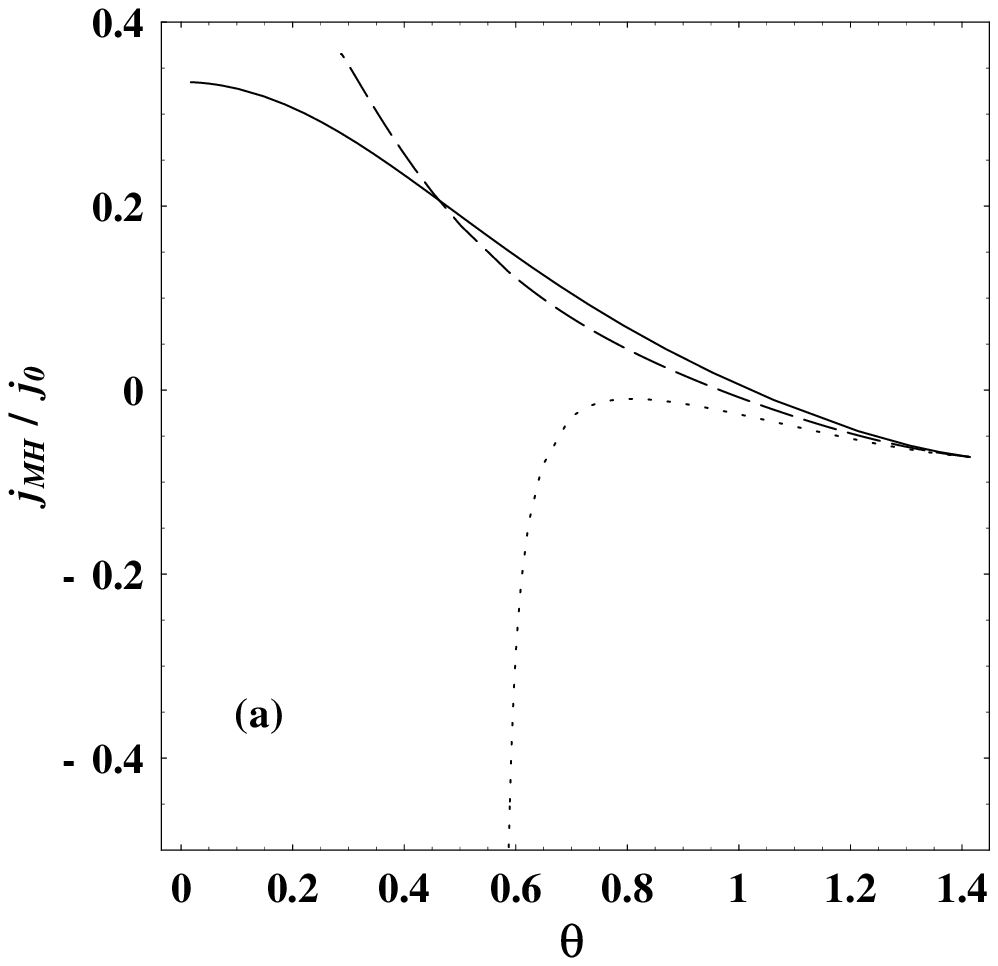}
\end{center}
\begin{center}
\plotone{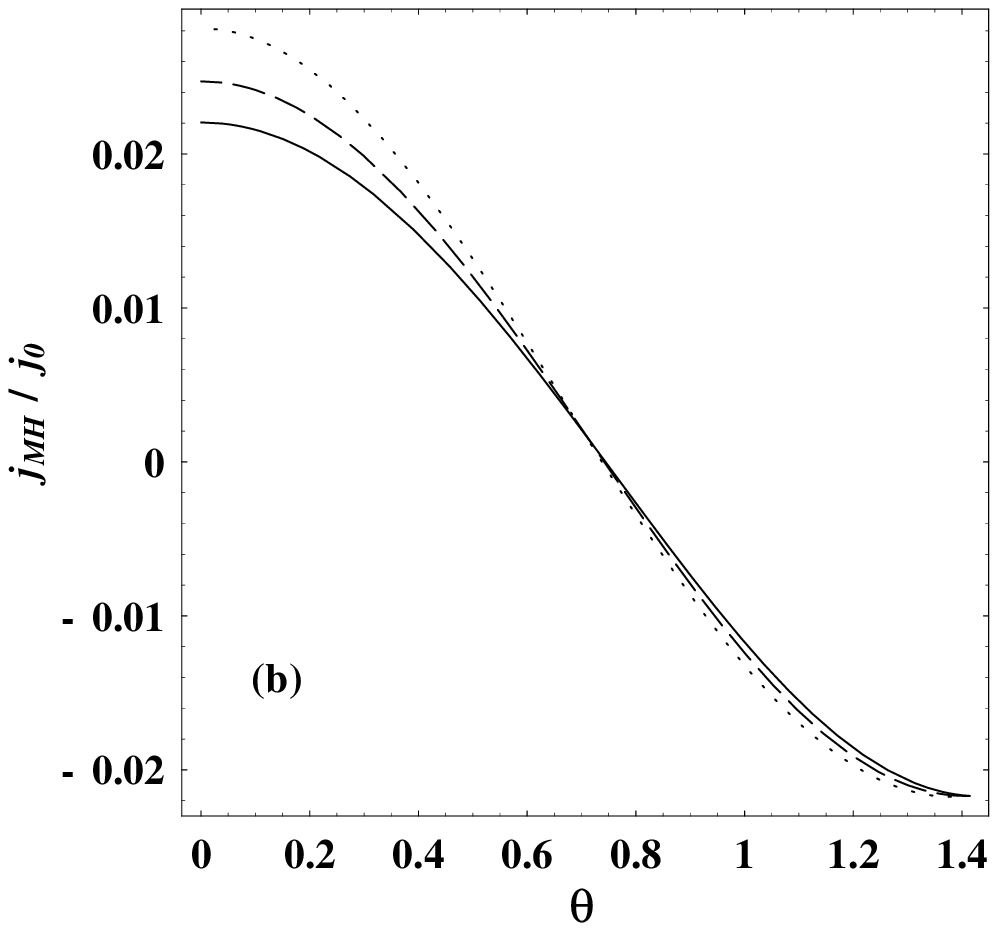}
\end{center}
\begin{center}
\plotone{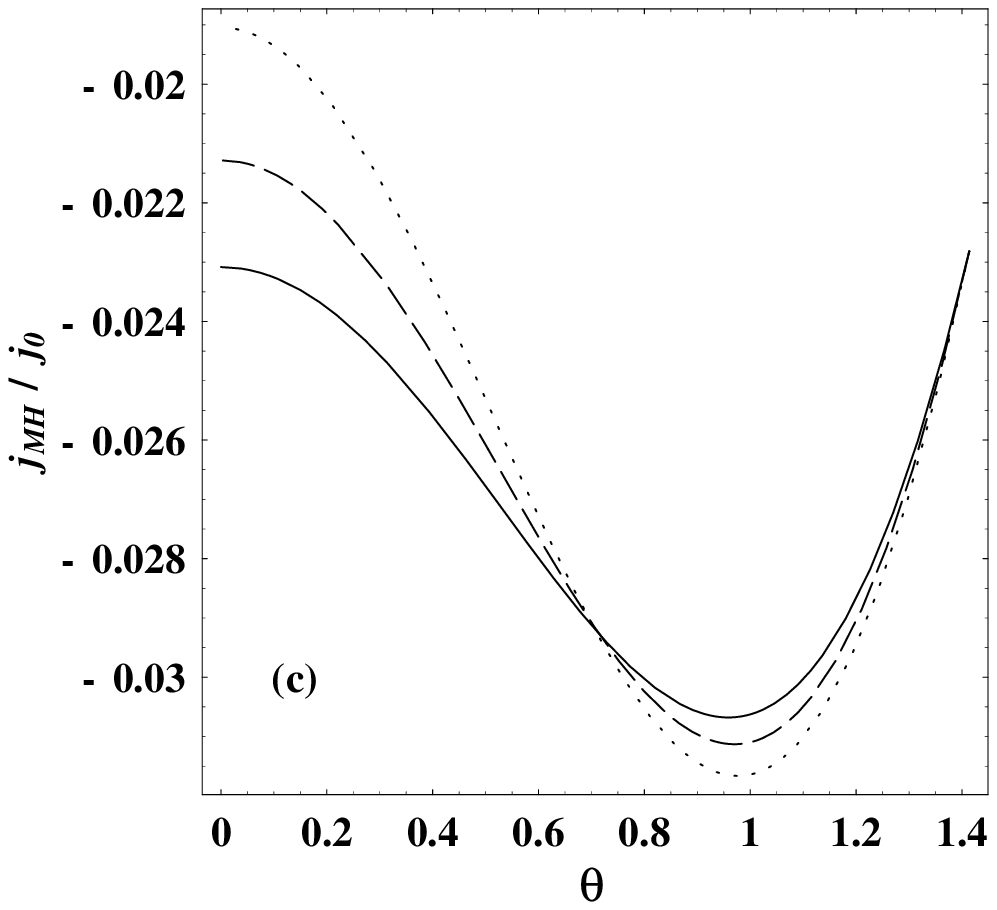}
\end{center}
\caption{The curves of $j_{MH}/j_0 $ versus $\theta $ from $\theta
_M $ to $\theta _L $ with $n = 3.0, 4.0 $ and $4.5$ in solid,
dashed and dotted lines, respectively. (a) $a_ * = 0.998$, (b) $a_
* = 0.3$, (c) $a_ * = 0.1$.}
\label{fig7}
\end{figure}

\clearpage
\begin{figure}
\epsscale{0.8}
\begin{center}
\plottwo{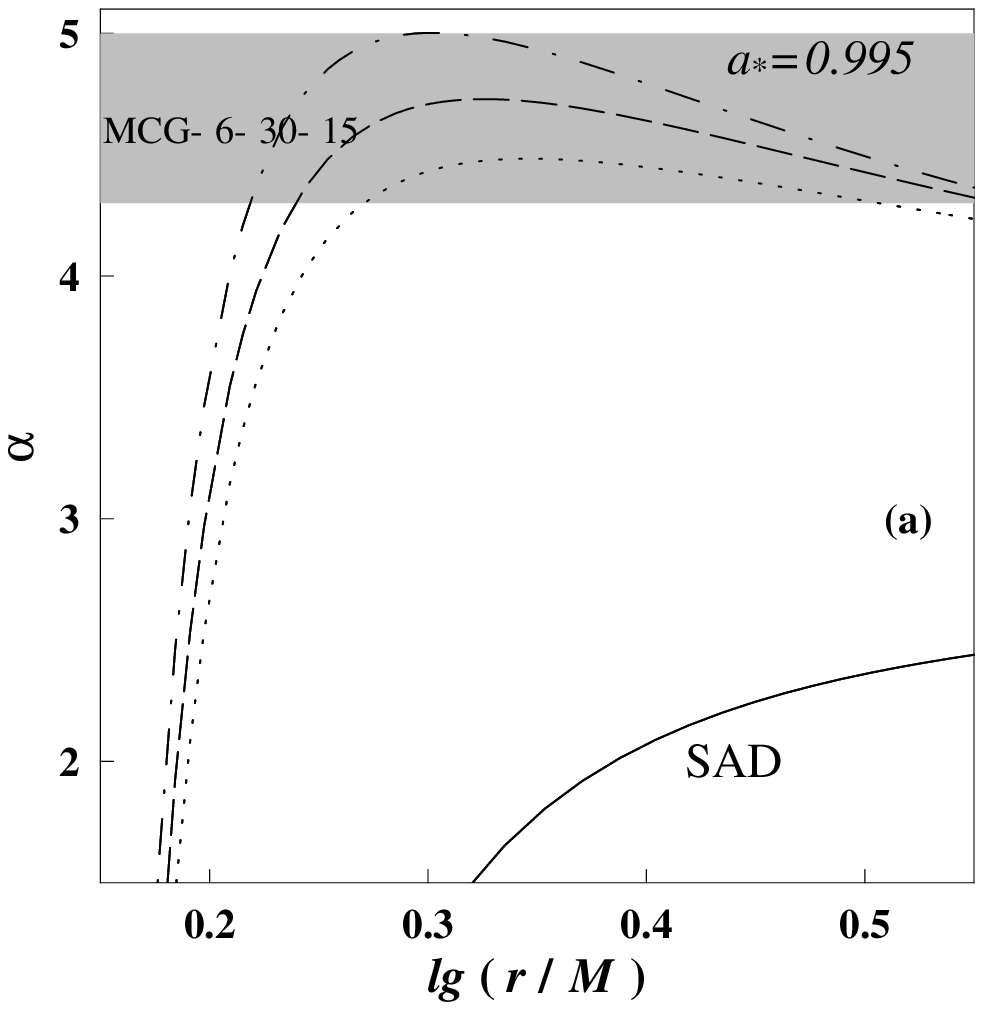}{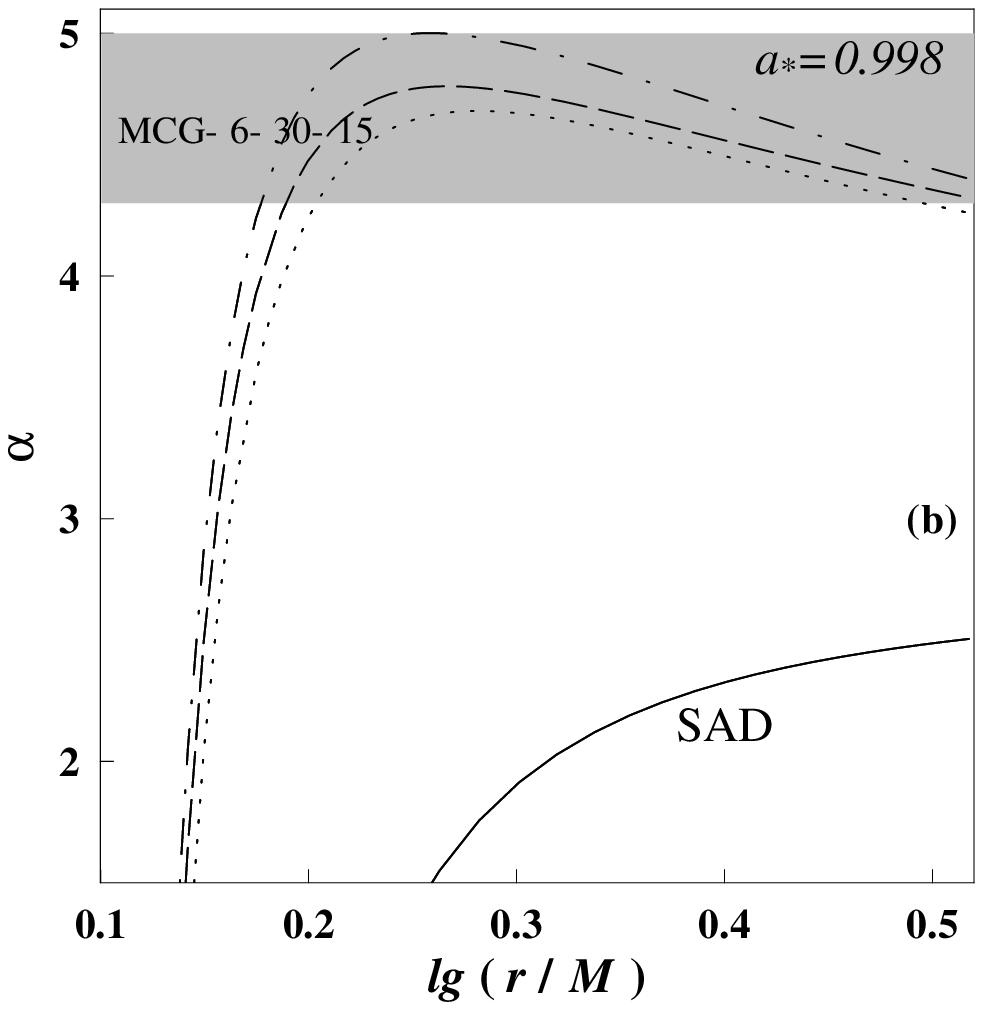}
\end{center}
\begin{center}
\plottwo{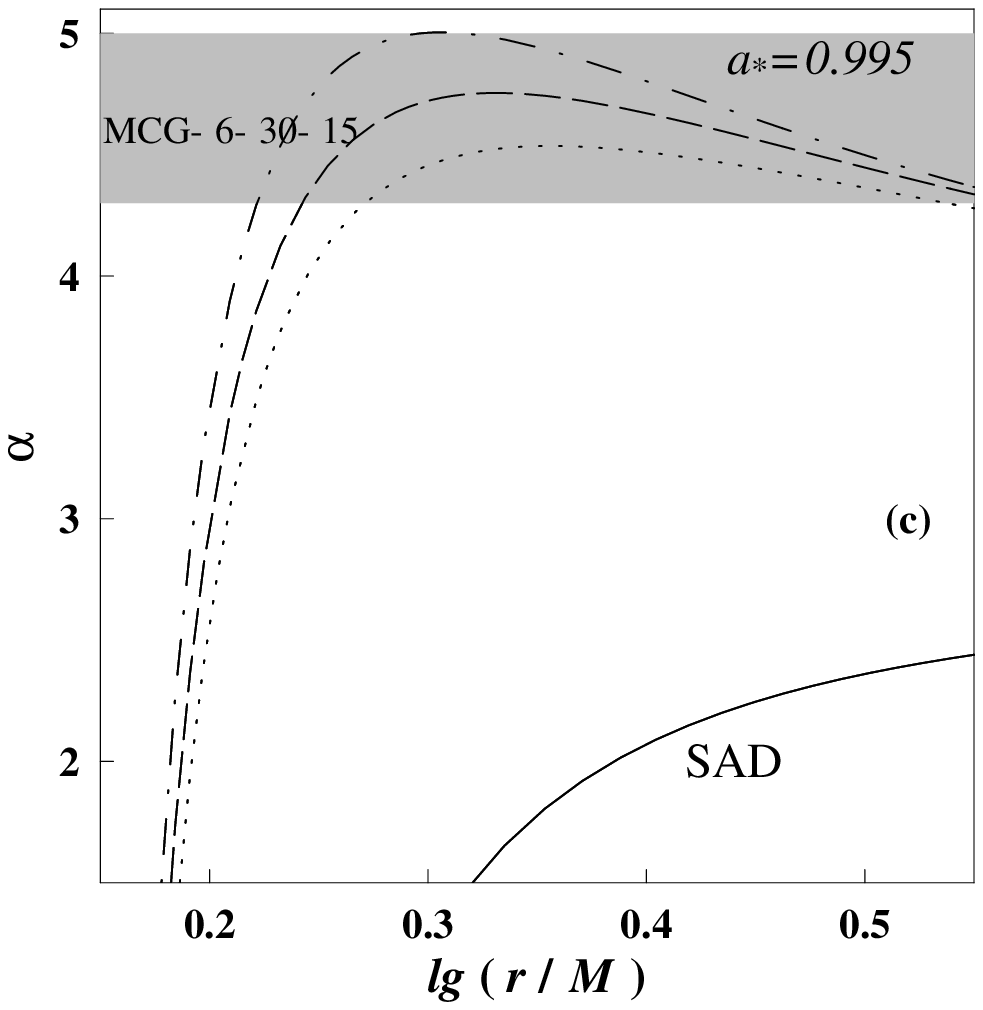}{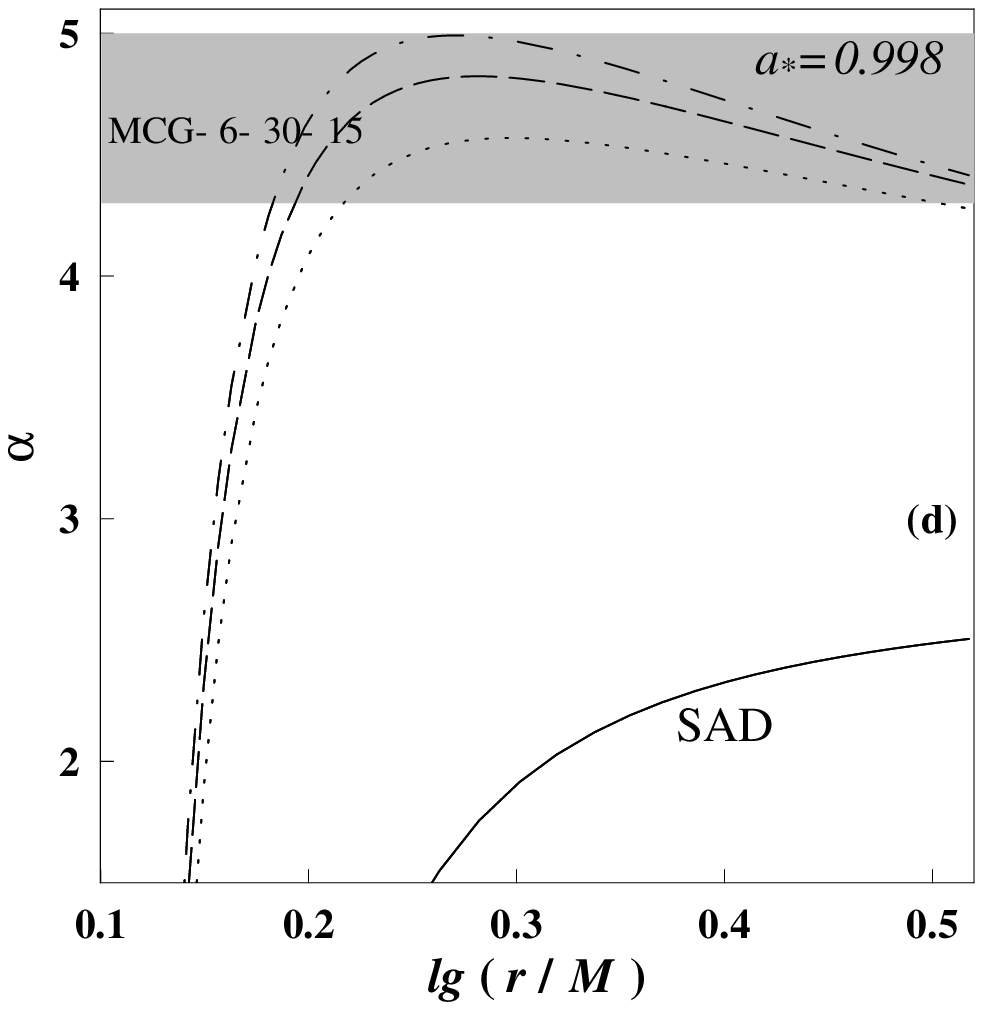}
\end{center}
\caption{The emissivity index of the Seyfert 1 galaxy MCG-6-30-15
inferred from the observation of \textit {XMM-Newton} is shown in
the shaded region. The emissivity index of a standard accretion
disk (SAD) is plotted in solid line. The emissivity index produced
by MC process is plotted versus $\lg(r/M)$  in dotted, dashed and
dot-dashed lines for $n = 6.0, 7.5$ and $9.2$ in (a), $n =
4.3,\mbox{ 6}.0$ and $7.3$ in (b), $n = 6.8, 8.0$ and $9.5$ in
(c), and $n = 5.8, 7.0$ and $7.8$ in (d), respectively. The
following parameters are imposed: $\theta _L = 0.40\pi $ in (a)
and (b), $\theta _L = 0.45\pi $ in (c) and (d); $a_
* = 0.995$ in (a) and (c), $a_ * = 0.998$ in (b)
and (d).}
\label{fig8}
\end{figure}


\clearpage
\begin{deluxetable}{crrrr}
\tabletypesize{\scriptsize} \tablecaption{Direction of
$\vec{S}_L^p$, $\vec{S}_E^p$ and corresponding quantities above
and below EQP of the BH (EQP in Table 1 and Table 2 below is an
abbreviation for equatorial plane)} \tablewidth{0pt}
\tablehead{\colhead{Position} & \colhead{$\vec{B}^p$} &
\colhead{$\vec{B}^T$} & \colhead{Current through $m$-loop} &
\colhead{$\vec{S}_L^p \quad and \quad \vec{S}_E^p$} } \startdata
\begin{tabular}{cc}
\begin{tabular}{c}
above \\ EQP
\end{tabular} &
\begin{tabular}{c} \\ OMCR \\ \\ \tableline \\ IMCR
\end{tabular}
\end{tabular} &
\begin{tabular}{c}
 From the BH \\ to the disk \end{tabular}&
\begin{tabular}{c} \\ Opposite to $\vec{m}$ \\ \\ \tableline \\ Same
as $\vec{m}$ \end{tabular} &
\begin{tabular}{c} \\ $I>0$, downwards \\ \\
\tableline \\  $I<0$, upwards \end{tabular} &
\begin{tabular}{c}
\\ From the BH to the disk \\ \\ \tableline \\ From the disk to the
BH \end{tabular}\\ \\

\tableline \\

\begin{tabular}{cc}
\begin{tabular}{c}
below \\ EQP
\end{tabular} &
\begin{tabular}{c} \\ OMCR \\ \\ \tableline \\ IMCR
\end{tabular}
\end{tabular} &
\begin{tabular}{c} From the disk  \\ to the BH  \end{tabular}&
\begin{tabular}{c} \\Same
as $\vec{m}$ \\  \\ \tableline \\ Opposite to $\vec{m}$
\end{tabular} &
\begin{tabular}{c} \\ $I<0$, upwards \\ \\
\tableline \\ $I>0$, downwards  \end{tabular} &
\begin{tabular}{c}
\\ From the BH to the disk \\ \\ \tableline \\ From the disk to the
BH \end{tabular} \\ \\
\enddata
\end{deluxetable}


\begin{deluxetable}{crrrr}
\tabletypesize{\scriptsize} \tablecaption{Direction of
$(\vec{S}_{E}^p)_1$ and corresponding quantities above and below
EQP of the BH } \tablewidth{0pt}
 \tablehead{\colhead{Case} &
\colhead{$\vec{v}^F$} & \colhead{$\vec{E}^p$} &
\colhead{$\vec{B}^T$} & \colhead{$(\vec{S}_E^p)_1$}}
\startdata
\begin{tabular}{ccc}
\begin{tabular}{c} above \\ EQP \end{tabular} &
\begin{tabular}{c} \\ A \\ \\ B \\ \\ C \end{tabular} &
\begin{tabular}{c} \\ $\Omega^F > \Omega^H > \omega$ \\ \\ $\Omega^H > \Omega^F > \omega$ \\ \\ $\Omega^H > \omega > \Omega^F$ \end{tabular}
 \end{tabular} &
 \begin{tabular}{c} \\  Same as $\vec{m}$ \\ \\ Same as $\vec{m}$\\ \\ Opposite to $\vec{m}$ \end{tabular}&
 \begin{tabular}{c} \\ \begin{tabular}{c} Outward and  normal to $\vec{B}^p$ \end{tabular} \\ \\ \begin{tabular}{c}  Outward and  normal to
 $\vec{B}^p$ \end{tabular} \\ \\ \begin{tabular}{c}  Inward and  normal to
 $\vec{B}^p$\end{tabular}
 \end{tabular}&
 \begin{tabular}{c} \\ Same as $\vec{m}$ \\ \\ Opposite to $\vec{m}$ \\ \\ Opposite to $\vec{m}$
 \end{tabular} &
 \begin{tabular}{c} \\ Opposite to $\vec{B}^p$ \\ \\ Same as  $\vec{B}^p$ \\ \\ Opposite to
 $\vec{B}^p$ \end{tabular} \\

\tableline \\

\begin{tabular}{ccc}
\begin{tabular}{c} below \\ EQP \end{tabular} &
\begin{tabular}{c}   D \\ \\ E \\ \\ F \end{tabular} &
\begin{tabular}{c}  $\Omega^F > \Omega^H > \omega$ \\ \\ $\Omega^H > \Omega^F > \omega$ \\ \\ $\Omega^H > \omega > \Omega^F$ \end{tabular}
 \end{tabular} &
 \begin{tabular}{c}  Same as $\vec{m}$ \\ \\ Same as $\vec{m}$\\ \\ Opposite to $\vec{m}$ \end{tabular}&
 \begin{tabular}{c}  \begin{tabular}{c} Outward and  normal to $\vec{B}^p$ \end{tabular} \\ \\ \begin{tabular}{c}  Outward and  normal to
 $\vec{B}^p$ \end{tabular} \\ \\ \begin{tabular}{c}  Inward and  normal to
 $\vec{B}^p$\end{tabular}
 \end{tabular}&
 \begin{tabular}{c}  Opposite to  $\vec{m}$ \\ \\ Same as $\vec{m}$ \\ \\ Same as $\vec{m}$
 \end{tabular} &
 \begin{tabular}{c} Same as $\vec{B}^p$ \\ \\ Opposite to $\vec{B}^p$ \\ \\ Same as $\vec{B}^p$ \end{tabular}
 \\
\enddata
\end{deluxetable}


\begin{deluxetable}{cccccc}
\tabletypesize{\scriptsize} \tablecaption{The parameters $a_*$,
$n$, and $\theta_L$ adapting the emissivity index to the
observations} \tablewidth{0pt} \tablehead{\colhead{$\theta_L$} &
\colhead{$a_*=0.990$} & \colhead{$a_*=0.992$} &
\colhead{$a_*=0.994$} & \colhead{$a_*=0.996$} &
\colhead{$a_*=0.998$}} \startdata
 $0.40\pi$ & $5.9-11.1$ & $5.6-10.4$ & $4.9 - 9.6$ & $3.8 -
 8.7$ & $ 4.0 - 7.3$ \\
 $0.45\pi$ & $6.4-11.3$ & $6.1-10.6$ & $5.8 - 9.9$ & $5.3 -
 9.0$ & $ 4.1 - 7.8$ \\
 $0.48\pi$ & $6.5-11.4$ & $6.3-10.7$ & $6.0 - 10.0$ & $5.6 -
 9.2$ & $ 4.9 - 8.0$ \\
\enddata
\end{deluxetable}


\begin{thebibliography}

\bibitem[1]{b1}{Bardeen J. M., Press W. H., and Teukolsky S. A., 1972, ApJ,
178, 347}

\bibitem[2]{b2}{Blandford R. D., 1976, MNRAS, 176, 465}

\bibitem[3]{b3}{Blandford R. D., Znajek R. L., 1977, MNRAS, 179, 433}

\bibitem[4]{b4}{Blandford R. D., 1999, in Sellwood J. A., Goodman J., eds, ASP Conf. Ser.
Vol. 160, Astrophysical Discs: An EC Summer School,Astron. Soc.
Pac., San Francisco, p.265}

\bibitem[5]{b11}{Ghosh P., Abramowicz M. A., 1997, MNRAS, 292, 887}

\bibitem[6]{b5}{Li L. -X. 2000, ApJ, 533, L115}

\bibitem[7]{b6}{Li L. -X. 2002, ApJ, 567, 463 (Li02a)}

\bibitem[8]{b7}{Li L. -X. 2002, A{\&}A, 392, 469 (Li02b)}

\bibitem[9]{b8}{Li L. -X. 2002, Phys. Rev. D, 65, 084047 (Li02c)}

\bibitem[10]{b8}{Li L. -X., Paczynski B., 2000, ApJ, 534, L 197}

\bibitem[11]{b9}{Macdonald D., Thorne K. S., 1982, MNRAS, 198, 345 (MT)}

\bibitem[12]{b10}{Wald R. M., 1984, {\it General Relativity}, Chicago Univ. Press, Chicago}

\bibitem[13]{b11}{Wang D. X., Xiao K., Lei W. H., 2002, MNRAS, 335, 655
(WXL)}

\bibitem[14]{b12}{Wang D. X., Lei W. H., Ma R. Y., 2003, MNRAS, (accepted,
WLM)}



\end{thebibliography}
\end{document}